\documentclass{llncs}
\pdfoutput=1
\usepackage{jason}
\usepackage{paralist}
\usepackage{url}
\usepackage{stmaryrd}
\usepackage{tikz}
\usetikzlibrary{arrows,automata,shapes}
\usetikzlibrary{backgrounds,fit,decorations.pathreplacing}

\renewcommand{\true}{1}
\renewcommand{\false}{0}
\newcommand{\match}{\true_{\rm T}}
\newcommand{\nomatch}{\false_{\rm T}}
\newcommand{\noreqmatch}{\bot_{\rm T}}
\newcommand{\allow}{\true_{\rm P}}
\newcommand{\deny}{\false_{\rm P}}
\newcommand{\na}{\bot_{\rm P}}

\newcommand{\tand}{\mathbin{\mathsf{and}_{\rm T}}}
\newcommand{\tor}{\mathbin{\mathsf{or}_{\rm T}}}
\newcommand{\tnot}{\mathop{\mathsf{not}_{\rm T}}}
\newcommand{\tnull}{\mathsf{null}_{\rm T}}
\newcommand{\tdecisions}{\mathsf{Dec}_{\rm T}}
\newcommand{\pand}{\mathbin{\mathsf{and}_{\rm P}}}

\newcommand{\pnot}{\mathop{\mathsf{not}_{\rm P}}}
\newcommand{\dbd}{\mathop{\mathsf{dbd}_{\rm P}}}
\newcommand{\abd}{\mathop{\mathsf{abd}}}
\newcommand{\andcup}{\mathbin{\mathsf{and}_{\cup}}}
\newcommand{\orcup}{\mathbin{\mathsf{or}_{\cup}}}
\newcommand{\andcap}{\mathbin{\mathsf{and}_{\cap}}}

\newcommand{\orcap}{\mathbin{\mathsf{or}_{\cap}}}
\newcommand{\pdecisions}{\mathsf{Dec}_{\rm P}}
\newcommand{\tweakening}{\mathop{\mathsf{opt}_{\rm T}}}
\newcommand{\kweakening}{\mathop{\sim}}
\newcommand{\sem}[1]{\ensuremath{\llbracket #1 \rrbracket}}
\newcommand{\tsem}[1]{\sem{#1}_{\rm T}}
\newcommand{\psem}[1]{\sem{#1}_{\rm P}}

\newcommand{\weakkland}{\sqcap}
\newcommand{\weakklor}{\sqcup}
\newcommand{\strongkland}{\mathbin{\tilde{\sqcap}}}
\newcommand{\strongklor}{\mathbin{\tilde{\sqcup}}}

\newcommand{\kleenesup}{\mathbin{\dot \sqcup}}

\newcommand{\ptackle}{PTaCL\xspace}

\renewenvironment{proof}[1][Proof]
{\begin{trivlist} \item \textbf{#1}\ \ }{\end{trivlist}}

\begin{document}

\title{\ptackle: A Language for Attribute-Based\\ Access Control in Open Systems}
\titlerunning{\ptackle: A Target-Based Authorization Language}

\author{Jason Crampton\inst{1} \and Charles Morisset\inst{1,2}}
%
%\authorrunning{An Auto-Delegation Mechanism for Access Control Systems}

\institute{
  Information Security Group,\\
  Royal Holloway, University of London, \\
  Egham, Surrey TW20 0EX, U.K. e-mail: \\
  {\tt Jason.Crampton@rhul.ac.uk}
 \and
 Security Group, \\
 Istituto di Informatica e Telematica (IIT), C.N.R., \\
 Via Giuseppe Moruzzi, 1, 56124 Pisa, Italy. e-mail: \\
 {\tt Charles.Morisset@iit.cnr.it}
}

\maketitle

\begin{abstract}
Many languages and algebras have been proposed in recent years for the specification of authorization policies.
For some proposals, such as XACML, the main motivation is to address real-world requirements, typically by providing a complex policy language with somewhat informal evaluation methods; others try to provide a greater degree of formality~--~particularly with respect to policy evaluation~--~but support far fewer features.
In short, there are very few proposals that combine a rich set of language features with a well-defined semantics, and even fewer that do this for authorization policies for attribute-based access control in open environments.
In this paper, we decompose the problem of policy specification into two distinct sub-languages: the policy target language (PTL) for target specification, which determines when a policy should be evaluated; and the policy composition language (PCL) for building more complex policies from existing ones.
We define syntax and semantics for two such languages and demonstrate that they can be both simple and expressive.
\ptackle, the language obtained by combining the features of these two sub-languages, supports the specification of a wide range of policies.
However, the power of \ptackle means that it is possible to define policies that could produce unexpected results.
We provide an analysis of how PTL should be restricted and how policies written in PCL should be evaluated to minimize the likelihood of undesirable results.
\end{abstract}

%\category{D.4.6}{Security and Protection}{Access Controls}
%A category including the fourth, optional field follows...
%\category{K.6.5}{Management of Computing and Information Systems}{Security and Protection}
%\category{H.2.3}{Languages}{Data manipulation languages (DML)}

%\terms{Design, Security, Theory}


\section{Introduction}

One of the fundamental security services in computer systems is \emph{access control}, a mechanism for constraining the interaction between (authenticated) users and protected resources.
Generally, access control is implemented by an  authorization service, which includes an \emph{authorization decision function} (ADF) for deciding whether a user request to access a resource (an ``access request'') should be permitted or not.
In its simplest form an authorization decision function either returns an $\mathsf{allow}$  or a $\mathsf{deny}$ decision.

Many access control models and systems are \emph{policy-based}, in the sense that a request for access to protected resources is evaluated with respect to a policy that defines which requests are authorized.
Many languages have been proposed for the specification of authorization policies, perhaps the best known being XACML~\cite{bert:auth01,dami:fine02,xacml2.0}.
However, it is generally acknowledged that XACML suffers from having poorly defined and counterintuitive semantics~\cite{li:acce09,ni:dalg09}.
More formal approaches have provided well-defined semantics and typically use ``policy operators'' to construct complex policies from simpler sub-policies~\cite{back:alge04,bona:alge02,wije:prop03}.
However, such approaches tend to support fewer ``features'' than XACML.

In a ``closed'' information system -- one in which all authorized users are known to the system -- it is possible to authenticate users of the system and to ascribe an identity to processes associated with those users.
Hence, access control decisions and the policies that inform those decisions can be based on user identifiers.

Increasingly, it is necessary to define authorization policies for ``open'' systems, where we must make access control decisions based on user attributes, rather than identities.
Hence, access request formats need to change from the user-centric subject-object-action triples of classical access control models~\cite{bell:secu76,harr:prot76}, although such request formats are still widely used in the specification of access control models and authorization policy languages~\cite{bert:auth01,bona:alge02,dami:fine02,xacml2.0,wije:prop03}.

An authorization policy is typically defined by a target, a set of child policies and a decision-combining algorithm.
The target, either implicitly or explicitly, identifies a set of requests.
The policy is said to be ``applicable'' if the access request belongs to (or ``matches'') the target.
If a policy is applicable, then its child policies are evaluated and the results returned by those child policies are combined using the decision-combining algorithm.

Informally, a policy may be regarded as a tree, in which the leaf nodes return a ``conclusive'' decision ($\mathsf{allow}$ or $\mathsf{deny}$).
If a request does not match the target of a leaf policy then the evaluation of that policy returns a ``not applicable'' decision.
Hence, the set of possible decisions is $3$-valued. % even in the case when we can always evaluate the applicability of a request to a policy.

However, it may be the case that it is not possible to evaluate request applicability: perhaps the simplest case arises when the request is malformed.
But once the request format is extended to accommodate attribute-based access control, the problem of evaluating the applicability of a request becomes even more acute.
In other words, the result of request applicability is not necessarily binary: in particular, we must include a value that represents that some error has occurred while trying to evaluate request applicability.
Naturally, extending the set of results that can be returned when evaluating request applicability means that we need to reconsider policy evaluation.

We believe that existing proposals for authorization policy languages suffer from at least one of the following problems:
  \begin{compactitem}
    \item no support for attribute-based requests (and hence attribute-based authorization policies);
    \item a lack of formality in the definition of target and policy evaluation, leading to ambiguity about the meaning of policies;
    \item a poor understanding of the way in which attribute-based requests, targets and policies interact.
  \end{compactitem}
Our main objective is to define a policy language that addresses the same problem space as XACML 3.0~\cite{XACML3} while retaining the formality of recent work on policy algebras~\cite{back:alge04,bona:alge02,BH08,CH10,wije:prop03}.
XACML (eXtensible access control markup language) is a standardized language: XACML 2.0 was ratified in 2005; XACML 3.0 will add support for attribute-based access control and policy administration.
More specifically, our objectives are:
  \begin{compactitem}
    \item to define a request format that is appropriate for attribute-based authorization policies;
    \item to define a syntax for specifying policy targets;
    \item to formally define an evaluation method for those targets that is sufficiently robust to withstand deliberate attempts to exploit the greater freedom provided by our request format;
    \item to define a syntax for policies, which makes use of the policy target language;
    \item to formally define an evaluation method for those policies that is able to handle errors in target evaluation gracefully and securely.
  \end{compactitem}

In this paper, we develop two distinct languages for completely defining authorization policies.
Roughly speaking, our goals are to combine support for the wide variety of policies that can be defined in more informal approaches such as XACML with the more formal semantics with which policy algebras are furnished.
Our policy target language (PTL) provides a syntax for specifying policy targets, while our policy composition language (PCL), provides a language for combining policies (that is, constructing policy trees).
Together, we call this \ptackle, read ``p-tackle'', to denote \emph{policy target and composition language}.
We also provide ``authorization policy semantics'', which enable us to ascribe a meaning to a policy for a given request.
That meaning is determined by the target semantics and the composition semantics.

The main contribution of this work is therefore the definition of \ptackle, which, although far simpler syntactically than XACML 2.0 and 3.0, can express any desired target or policy, thanks to the functional completeness of PTL and PCL.
We specify precisely how to evaluate any target and policy expressed in \ptackle, thus providing the basis for a low-level language into which XACML policies, for example, could be compiled and evaluated.
Moreover, we identify the problem of {\em attribute-hiding attacks}, where a user deliberately suppresses attributes in order to gain favorable authorization decisions, and we propose different restrictions on the definition of a target in order to avoid such attacks.  
We note that such attacks are not peculiar to \ptackle; they are a potential problem for any attribute-based access control mechanism.  
We believe we are the first to identify and, therefore, propose mitigation strategies for, this type of attack.

In the next section, we define our request format and illustrate some of the challenges introduced by attribute-based access control.
Then, in Section~\ref{sec:target}, we define the syntax and evaluation method for targets.
In Section~\ref{sec:policies}, we define policy syntax and evaluation.
In this section, we reflect on the problems that might arise because of the more flexible request format we use and explain how those problems inform the development of \ptackle.
We also explain how PTL can be restricted to provide certain guarantees about the decisions returned by policy evaluation, thereby addressing the problem attribute-hiding attacks.
We conclude the paper with a discussion of related work and some ideas for future work.

\section{Attribute-Based Requests}
\label{sec:overview}

The simplest authorization policy languages assume that an access request comprises three identifiers: the requester, the resource to which access is requested, and the type of the requested interaction (such as read, write, etc), often known as subject, object and action, respectively.
The authorization decision function (ADF) associated with a given language will take that request and an authorization policy as input and return a decision.
For more complex languages, the ADF may require additional information, such as the roles or security groups associated with a user, in order to make a decision.
These attributes may be ``pushed'' with the request or ``pulled'' from authoritative information sources (such as the policy information points in the XACML architecture).
The increasingly ``open'' nature of distributed computer systems, where the user population is not known in advance, requires authorization languages that are not based on user identities.
For this reason, attribute-based access control (ABAC) and languages that support ABAC are expected to become increasingly important.

\ptackle comprises two sub-languages: PTL for target specification and PCL for policy specification.
Policies written in \ptackle are used to evaluate access requests that may contain arbitrary attributes associated with users, resources and actions.

We model a request as a set of name-value pairs, where each name specifies an attribute and each value specifies a value for the corresponding attribute.
In the simplest situation, for example, we might have attribute names such as \texttt{subject}, \texttt{object} and \texttt{action}, and a request might have the form
 \[%begin{equation}\label{eq:standard-request-format}
   \set{(\mbox{\texttt{subject}},\mathit{alice}),(\mbox{\texttt{object}},\mathit{test.txt}),(\mbox{\texttt{action}},\mathit{read})}.
 \]%end{equation}
The above request is no different from the usual view of an access request as a subject-object-action triple.
However, the request format described above is not limited to requests of this form and can be used to represent requests that do not contain identifiers for subjects, objects and actions.
We could, for example, have a request of the form
 \[%begin{equation}\label{eq:non-standard-request-format}
   \set{(\mbox{\texttt{role}},\mathit{nurse}),(\mbox{\texttt{object}},\mathit{test.txt}),(\mbox{\texttt{action}},\mathit{read})}.
 \]%end{equation}
An attribute name may appear multiple times in the request; the above request could include multiple role identifiers, for example.
The use of some set of name-value pairs, rather than the fixed format subject-object-action triples (as used in XACML 2.0~\cite{xacml2.0} and most other policy languages), means that we can specify targets and requests with greater freedom than is usually the case.
However, the greater freedom with which requests can be specified also means that we have to take greater care in the specification of policies.

As an example, we consider a simplified instance of the Chinese Wall policy, where a company $A$ defines a policy to protect a set of confidential resources.
Informally, this policy states that if a user is working for $A$, then she can access the (confidential) resource $o$, unless she is also working for $B$, the direct competitor of $A$, in which case the access is denied.
We consider the following requests:
  \begin{align*}
    r_1 &= \set{({\tt employer}, A), ({\tt confidential}, true)}; \\
    r_2 &= \set{({\tt employer}, A),({\tt employer}, B), ({\tt confidential}, true)}; \\
    r_3 &= \set{({\tt confidential}, false)}; \\
    r_4 &= \set{({\tt confidential}, true)}.
  \end{align*}

Informally, an ABAC policy defines a set of atomic policies (or rules), where each atomic policy describes the subset of requests to which it applies~--~the policy's \emph{target}~--~and the decision to take when it is applicable.
When a request does not belong to the policy's target, then this policy is non-applicable, which has a different meaning from saying that the request is denied.
The decisions returned by the evaluation of the atomic policies are then combined together using decision combination operators.

For instance, the policy enforced by the company $A$ should comprise two rules, the second of which is applicable to all requests and returns allow.
The first rule is applicable if the request contains ({\tt confidential}, $true$), and in this case, if the user works for $A$, then it is allowed, unless she also works for $B$, in which case it is denied.
The two rules are combined using a deny-overrides combination operator.
The first rule would not be applicable to request $r_3$ and hence the request would be allowed.
The first rule would be applicable to the remaining requests.
Therefore, the evaluation of $r_1$ would return allow, while the evaluation of request $r_2$ would return deny.

Note that if the user is able to {\em suppress} the element $({\tt employer}, B)$ in $r_2$, then the resulting request would be allowed.
We call such a situation a {\em partial attribute hiding attack}, where, by hiding some of her attributes, a user is able to obtain a more favorable authorization decision.
A second possibility is for the user to suppress all the employer attributes.
Hence, we might wish to insist that if the resource is confidential, then the request {\em must} contain information about the employer(s) of the requesting user, otherwise the evaluation of the request should fail.
In particular, $r_4$ must not be allowed, returning either deny or some appropriate evaluation-error decision.

We now describe \ptackle, which provides mechanisms to tackle the issues raised by this simple example, in particular by considering attribute requests instead of subject-object-action requests; by distinguishing between optional and mandatory attributes; and by stating two properties of monotonicity, thus allowing the detection of policies vulnerable to partial attribute hiding attacks.

\section{Targets}
\label{sec:target}
We first define a syntax for targets.
Then, in Section~\ref{sec:evaluation}, we will define how to evaluate a target with respect to a request.
We define three types of \emph{atomic target}:
  \begin{compactitem}
    \item $\tnull$ is a target;
    \item $n$ is a target, where $n$ is an attribute name;
    \item $(n,v,f)$ is a target, where $n$ is an attribute name, $v$ is an attribute value and $f$ is a binary predicate.
  \end{compactitem}
The most usual predicate is likely to be a test for (string) equality, but other predicates, such as $\leqslant$, $<$, $\geqslant$ and $>$, are possible.
For ease of exposition, we assume throughout that all attributes are of type string and that $f$ is string equality; henceforth we omit $f$ from the definition of an atomic target.

We build more complex targets by defining two binary target operators, $\tand$ and $\tor$, and two unary target operators, $\tweakening$ and $\tnot$.
Let $t$, $t_1$ and $t_2$ be targets.
Then the following terms are also targets:
  \[
    \tweakening t,\quad \tnot t,\quad (t_1 \tand t_2)\quad\mbox{and}\quad(t_1 \tor t_2).
  \]

The operators $\tweakening$ and $\tnot$ bind more tightly than $\tand $ and $\tor$: $\tweakening t\,\tand\,t'$, for example, is interpreted as $(\tweakening t)\,\tand\,t'$, rather than $\tweakening (t\,\tand\, t')$.
As we will see in Section~\ref{sec:evaluation}, the semantics of $\tor$ and $\tand$ are provided by associative, commutative binary operators on $\tdecisions$, so we can (and will) omit brackets from expressions of the form $(t_1 \tor (t_2 \tor \dots \tor t_k))$ and $(t_1 \tand (t_2 \tand \dots \tand t_k))$.

In Section~\ref{sec:policies}, we will define similar operators for policies and use a subscript P to distinguish them from target operators.
When no ambiguity can occur we will omit the subscripts T and P.
\renewcommand{\tand}{\mathbin{\mathsf{and}}}%
\renewcommand{\tor}{\mathbin{\mathsf{or}}}%
\renewcommand{\tnot}{\mathop{\mathsf{not}}}%
\renewcommand{\tweakening}{\mathop{\mathsf{opt}}}%
\renewcommand{\tnull}{\mathsf{null}}%
%

%, which respectively correspond to the negation and to the operator making a target ``optional'', as $\weakening t$ evaluates to $\nomatch$ if $t$ evaluates to $\noreqmatch$.
%We will use these operators to define the evaluation of more complex targets.

%  \begin{align*}
%    [[t_1 \weakkland t_2]]_q &=
%      \begin{cases}
%        \noreqmatch & \text{if $[[t_1]]_q = \noreqmatch$ or $[[t_2]]_q = \noreqmatch$} \\
%        [[t_1]]_q \wedge [[t_1]]_q & \text{otherwise}
%      \end{cases} \\
%    [[t_1 \weakklor t_2]]_q &=
%      \begin{cases}
%        \match & \text{if $[[t_1]]_q = \match$ or $[[t_2]]_q = \match$} \\
%        \nomatch & \text{if $[[t_1]]_q = [[t_2]]_q = \nomatch$} \\
%        \noreqmatch & \text{otherwise}
%      \end{cases}
%  \end{align*}

\subsection{Evaluation}
\label{sec:evaluation}

A target is evaluated with respect to a request, represented as a set of name-value pairs (as described in Section~\ref{sec:overview}).
Informally, a request is said to ``match'' an atomic target if the name of one of the attribute pairs in the request is the same as the name defined in the atomic target and the predicate $f$ evaluated at $v$ and the corresponding value in the request is true.
If no such pair exists in the request, then the request does not match the target.

The ``universal'' target $\tnull$ is matched by all requests; the target $n$ is matched by all requests that include an attribute pair $(n,v)$ for any value $v$; the target $(n,v)$ is matched by any request that includes the specific attribute pair $(n,v)$.
The target \texttt{employer}, for example, is matched by requests $r_1$ and $r_2$ defined in Section~\ref{sec:overview} but not by the requests $r_3$ and $r_4$.

% ~\eqref{eq:non-standard-request-format}, but not by request~\eqref{eq:standard-request-format}; the target $(\mbox{\texttt{role}},\mathit{doctor})$ is not matched by either request.

In addition, we may wish to distinguish the case where the request does not include the attribute name at all from the case where the attribute name was found, but with a value that does not match.
Consider the atomic target $(\mbox{\texttt{employer}},B)$:
then request $r_1$ has a matching attribute name (\texttt{employer}), but $A \ne B$;
in contrast, requests $r_3$ and $r_4$ do not include any matching attribute.
% to insist that a request includes a particular attribute name, so we raise an alarm if a request does not match a target that specifies a mandatory attribute name.

Informally, a request must match both $t_1$ and $t_2$ for it to match target $(t_1 \tand t_2)$, while a request is only required to match one of $t_1$ and $t_2$ for it to match target $(t_1 \tor t_2)$.
By default, a request is required to match a target $t$; we can relax this requirement, while retaining the possibility of matching $t$, by writing $\tweakening t$.

More formally, we define the set of target evaluation decisions $\tdecisions$ to be $\set{\match,\nomatch,\noreqmatch}$,%
\footnote{We will use analogous notation for decisions, where $\allow$ will denote an ``allow'' decision and $\deny$ will denote a ``deny'' decision.}
where $\noreqmatch$ denotes that a request does not include the attribute name, $\match$ denotes that a request matches an atomic target, and $\nomatch$ denotes that a request includes the attribute name but the predicate doesn't hold.%

We define the binary operators $\weakkland$, $\weakklor$, $\strongkland$ and $\strongklor$ on $\set{\match,\nomatch,\noreqmatch}$ in Fig.~\ref{fig:applicability-logical-operators}.
These operators correspond to the weak and strong Kleene operators~\cite{kleene:intr50}, respectively.
We also define two unary operators $\lnot$ and $\kweakening$ in Fig.~\ref{fig:applicability-logical-operators}.
Finally, we define the total order $\match > \nomatch > \noreqmatch$ on $\tdecisions$ and let $\kleenesup$ denote the least upper bound operator on this ordered set.

\begin{figure*}[t]
  \subfigure[]{
    \begin{minipage}{.185\textwidth}\small
      \[
        \begin{array}{c|ccc}
          \weakkland & \match & \nomatch & \noreqmatch \\
        \hline
          \match & \match & \nomatch & \noreqmatch \\
          \nomatch & \nomatch & \nomatch & \noreqmatch \\
          \noreqmatch & \noreqmatch & \noreqmatch & \noreqmatch \\
        \end{array}
      \]
    \end{minipage}}
  \subfigure[]{
    \begin{minipage}{.185\textwidth}\small
      \[
        \begin{array}{c|ccc}
          \weakklor & \match & \nomatch & \noreqmatch \\
        \hline
          \match & \match & \match & \noreqmatch \\
          \nomatch & \match & \nomatch & \noreqmatch \\
          \noreqmatch & \noreqmatch & \noreqmatch & \noreqmatch \\
        \end{array}
      \]
    \end{minipage}}
  \subfigure[]{
    \begin{minipage}{.185\textwidth}
      \[
        \begin{array}{c|ccc}\small
          \strongkland & \match & \nomatch & \noreqmatch \\
        \hline
          \match & \match & \nomatch & \noreqmatch \\
          \nomatch & \nomatch & \nomatch & \nomatch \\
          \noreqmatch & \noreqmatch & \nomatch & \noreqmatch \\
        \end{array}
      \]
    \end{minipage}}
  \subfigure[]{
    \begin{minipage}{.185\textwidth}\small
      \[
        \begin{array}{c|ccc}
          \strongklor & \match & \nomatch & \noreqmatch \\
        \hline
          \match & \match & \match & \match \\
          \nomatch & \match & \nomatch & \noreqmatch \\
          \noreqmatch & \match & \noreqmatch & \noreqmatch \\
        \end{array}
      \]
    \end{minipage}}
    \subfigure[]{
      \begin{minipage}{.185\textwidth}\small
        \[
          \begin{array}{c|c|c}
            X & \lnot X & \kweakening X \\ \hline
            \match & \nomatch & \match \\
            \nomatch & \match & \nomatch \\
            \noreqmatch & \noreqmatch & \nomatch \\
          \end{array}
        \]
      \end{minipage}}
%    \subfigure[Optional]{
%      \begin{minipage}{.1853\textwidth}
%        \[
%          \begin{array}{c|c}
%            X & \kweakening X \\
%          \hline
%            \match & \match \\
%            \nomatch & \nomatch \\
%            \noreqmatch & \nomatch \\
%          \end{array}
%        \]
%      \end{minipage}}
\caption{Binary and unary operators on the target decision set $\set{\match,\nomatch,\noreqmatch}$}\label{fig:applicability-logical-operators}
\end{figure*}

%CM I put the two following operators as a separated figure for now, but we can merge it later with the previous Figure
% \begin{figure}
%   \subfigure[Negation]{
%     \begin{minipage}{.45\textwidth}
%       \[
%         \begin{array}{c|c}
%           X & \lnot X  \\ \hline
%           \match & \nomatch \\
%           \nomatch & \match  \\
%           \noreqmatch & \noreqmatch  \\
%         \end{array}
%       \]
%     \end{minipage}}
%   \subfigure[Optional]{
%     \begin{minipage}{.45\textwidth}
%       \[
%         \begin{array}{c|c}
%           X & \kweakening X \\
%         \hline
%           \match & \match \\
%           \nomatch & \nomatch \\
%           \noreqmatch & \nomatch \\
%         \end{array}
%       \]
%     \end{minipage}}
%     \caption{Two unary operators on the target decision set $\set{\match,\nomatch,\noreqmatch}$}\label{fig:applicability-unary-operators}
% \end{figure}

Given a request $q$, we write $\tsem{t}(q)$ to denote the evaluation of $t$ with respect to $q$.
That is, $\tsem{t}(q) \in \tdecisions$.
As for target operators, we will omit the subscript T where no ambiguity can arise.
First, we define, for all requests $q$ and for all attributes $n$ and all values $v$,
  \[
    \sem{\tnull}(q) = \match \qquad\text{and}\qquad
    \sem{n}(\emptyset) = \sem{(n,v)}(\emptyset) = \noreqmatch.
  \]
We then define the evaluation of targets $n$ and $(n,v)$ recursively.
  \begin{align*}
%    \sem{n}(\emptyset) = \noreqmatch \qquad\text{and}\qquad
    \sem{n}(\set{(n',v')} \cup q) &=
      \begin{cases}
        \match & \text{if $n = n'$} \\
        \sem{n}(q)  & \text{otherwise}.
      \end{cases}
  \end{align*}
  \begin{align*}
    \sem{(n,v)}(\set{(n',v')} \cup q) &=
      \begin{cases}
        \match & \text{if $n = n'$, $v = v'$}\\
        \nomatch \kleenesup \sem{(n,v)}(q) &  \text{if $n = n'$, $v \ne v'$} \\
         \sem{(n,v)}(q)  & \text{otherwise}. \\
      \end{cases}
  \end{align*}
%where is the supremum (least upper bound) operator over the totally ordered set $\match > \nomatch > \noreqmatch$.
Note that, for all $q$, $\sem{n}(q)$ is either $\match$ or $\noreqmatch$.
In evaluating $(n,v)$, we compare each element of the request with the atomic target and do one of the following: we return $\match$ if a match is found; if the attribute name matches but the predicate doesn't hold then we record the fact that the attribute name matched and continue processing; otherwise, we simply continue processing.

Since $\kleenesup$ is a supremum operator, it is commutative and associative and hence can be applied to any subset of $\tdecisions$ without ambiguity.
Hence, for a non-empty request $q = \set{(n_1,v_1),\dots(n_k,v_k)}$, it is easy to see that we have
  \begin{align*}
    \sem{n}(q) &= \kleenesup\set{\sem{n}(\set{(n_i,v_i)}) : 1 \leqslant i \leqslant k}; \\
    \sem{(n,v)}(q) &= \kleenesup\set{\sem{(n,v)}(\set{(n_i,v_i)}) : 1 \leqslant i \leqslant k}.
  \end{align*}
In other words, we can evaluate the applicability of a request with respect to a target by splitting the request into single name-value pairs and evaluating each of these requests separately.
This, in turn, suggests that the evaluation of requests can be parallelized, with different TEFs specialized for the evaluation of requests for particular attribute names.

We then define the semantics of $\tnot t$, $\tweakening t$, $t_1 \tand t_2$ and $t_2 \tor t_2$ as follows:
  \begin{alignat*}{3}
    \sem{\tnot t}(q) &= \neg \sem{t}(q) & \quad  \sem{t_1 \tand t_2}&(q)~ &= \sem{t_1}(q) \weakkland \sem{t_2}(q) \\
    \sem{\tweakening t}(q) &= \kweakening\sem{t}(q)  \quad & \sem{t_1 \tor t_2}&(q)~ &= \sem{t_1}(q) \strongklor \sem{t_2}(q)
  \end{alignat*}

Here we see that $\tweakening$ ``weakens'' the target $t$ by converting a $\noreqmatch$ decision (missing attribute) into a $\nomatch$ decision (attribute not matched).
The target $\tweakening \mathtt{role}$, for example, evaluates to $\match$ if a request contains a role attribute pair and evaluates to $\nomatch$ (rather than $\noreqmatch$) if no such pair is present in the request.

It is important to note that the semantics for the $\tand$ operator are provided by weak conjunction $\weakkland$, not by $\strongkland$.
The point here is that a target is specified as part of a policy and it should not be possible to force target evaluation to return $\nomatch$ when the target is a conjunction and at least one of the conjuncts is mandatory.
(Had we combined targets using $\strongkland$, if $t_1$ were to evaluate to $\nomatch$ and $t_2$ were to evaluate to $\noreqmatch$, then $t_1 \strongkland t_2$ would evaluate to $\nomatch$, not the desired $\noreqmatch$.)

\subsection{Interface targets}

An atomic target of the form $(n,v)$ requires that a particular attribute value must appear in a request (to obtain a match).
Such targets are little different conceptually from those defined in XACML 2.0 and other authorization languages and are, therefore, of limited novelty or interest here.\footnote{Targets in XACML 2.0 only consider subjects, objects and actions; targets in the draft XACML 3.0 do consider other types of attributes.}

In contrast, targets of the form $n$, have not previously been seen in the literature on authorization languages (to the best of our knowledge).
A target of the form $n$ can be used to define a target that enforces a ``request interface'': a target of the form
  \[ \tweakening(n_1 \tand n_2 \tand \dots \tand n_k), \]
for example, only matches a request that contains particular named attributes (corresponding to $n_1,\dots,n_k$); the evaluation of a request that doesn't contain all the required attributes will evaluate to $\nomatch$ (because of the $\tweakening$).
In this way, we can construct a target that ``guards'' conventional subject-object-action policies and others that can respond to requests containing other types of attributes.

More complex ``mixed'' interfaces can also be constructed.
An access control list is a type of access control data structure that is widely used in operating systems.
The target for a policy used to represent an access control list for object $\mathit{test.txt}$ would have the form
  \[
    \tweakening((\mbox{\texttt{object}},\mathit{test.txt}) \tand \mbox{\texttt{subject}} \tand \mbox{\texttt{action}}),
  \]
so that only requests that specify the desired object as well as including some subject and action would match.

% \subsection{Relational targets}

% It is worth noting that an atomic target cannot compare the values of two different attributes, which can be limiting in some cases. 
% For instance, consider the UNIX policy, where each file partitions the set of users into three subsets: the owner, the users belonging to the same group than the object, and the rest of the users. 

% Intuitively, we want the target of such a policy to establish a relationship between the attributes of the resource and those of the user requesting an access over this resource. 

% However, assuming that the request 
% Clearly, there are different ways to define such a policy, according to the attributes present in the request. 

% A first approach consists in requiring the attribute provider to also compute the required relationships between the different entities. 

\subsection{Target equivalence}

We say that two targets $t_1$ and $t_2$ are \emph{equivalent} if, for all requests $q$, $\sem{t_1}(q) = \sem{t_2}(q)$, and write $\sem{t_1} = \sem{t_2}$ to denote that $t_1$ and $t_2$ are equivalent.
We note the following properties of our target operators.

\begin{Pro}
For all targets $t$ and $t'$, we have
  \begin{alignat*}{3}
    \sem{\tnot (\tnot t)} &= \sem{t} &\quad& &\sem{\tweakening(t \tand t')} &= \sem{(\tweakening t) \tand (\tweakening t')} \\
    \sem{\tweakening(\tweakening t)} &= \sem{\tweakening t} &\quad& &\sem{\tweakening(t \tor t')} &= \sem{(\tweakening t) \tor (\tweakening t')}
  \end{alignat*}
\end{Pro}

\begin{proof}
All the above results can be established by considering suitable ``truth'' tables.
\end{proof}

\noindent%
Note, however, that
    \begin{align*}
      \sem{\tweakening(\tnot t)} &\ne \sem{\tnot(\tweakening t)} \\
      \sem{\tnot (t_1 \tand t_2)} &\ne \sem{(\tnot t_1) \tor (\tnot t_2)} \\
      \sem{\tnot(t_1\tor t_2)} &\ne \sem{(\tnot t_1) \tand (\tnot t_2)}
    \end{align*}
because we use weak conjunction and strong disjunction to provide the semantics for $\tand$ and $\tor$ respectively.
% %In the next section we prove that we can represent targets $\tnot(t_1 \tand t_2)$ in terms of $\tnot$, $\tweakening$, $\tand$ and $\tor$.

\subsection{On functional completeness}
\label{sec:targetcompleteness}

By way of motivation, we first observe that it might be useful to be able to define ``conditional'' interface targets, where the presence of one attribute in a request requires the presence of some other attribute.
Suppose, for example, we have two attribute names $n_1$ and $n_2$.
If a request doesn't contain attribute $n_1$ then the evaluation of the target should be $\nomatch$.
If, however, a request does contain $n_1$ then it must contain $n_2$.
In other words, we have the following ``match table'', where the row headers indicate the values taken by the evaluation of $n_1$ and the column headers indicate the values taken by $n_2$.
  \[
    \begin{array}{c|cc}
        & \match & \noreqmatch \\
    \hline
      \match & \match & \noreqmatch \\
      \noreqmatch & \nomatch & \nomatch \\
    \end{array}
  \]
By inspection of the match tables in Fig.~\ref{fig:applicability-logical-operators}, we see that the above table could be represented by the target $\kweakening x \strongkland y$, where $x$ and $y$ denote the evaluation of $n_1$ and $n_2$, respectively.
However, the semantics of $\tand$ are given by the operator $\weakkland$.
Hence, it would be useful to demonstrate that our chosen target operators $\tweakening$, $\tnot$, $\tor$ and $\tand$ are functionally complete.
In particular, we would prefer to define the interface target described above in terms of our existing operators, rather than having to introduce another type of target conjunction.
%
%Suppose the attribute name \texttt{object-auth-role} is an attribute associated with an object whose value is a role name that is authorized to access the object.
%Then we might define a target
%  \[
%    (\tweakening\mbox{\texttt{object-auth-role}}) \tand \mbox{\texttt{role}}
%  \]
%that requires a request then it should also contain at least one attribute that specifies a role

%We now prove that the logic $(\set{\match, \nomatch, \noreqmatch}, \strongklor, \strongkland, \lnot, \kweakening)$ is functionally complete, meaning that for all $n$ and any function $f: \tdecisions^n \rightarrow \tdecisions$, $f$ can be constructed using the operators $\strongkland, \strongklor, \lnot$ and $\kweakening$.
%Hence we may consider the operators $\weakklor$ and $\weakkland$ to be nothing more than syntactic sugar.
We now prove that for all $n$ and any function $f: \tdecisions^n \rightarrow \tdecisions$, $f$ can be constructed using the constants $\match$, $\nomatch$ and $\noreqmatch$ and the operators $\tweakening$, $\tnot$ and $\tor$.
We obtain this property by proving that the three-valued logic expressed over the set $\set{\nomatch,\match,\noreqmatch}$ and defined by the operators $\strongklor$, $\neg$ and $\kweakening$ is functionally complete, re-using a result of Jobe~\cite{Jobe62}, stated below.

\newtheorem{Theorem}{Theorem}

\begin{Theorem}[Jobe 1962]
The three-valued logic $E$ expressed over the set $\set{1, 2, 3}$ and defined by the operators $\bullet, E_1$ and $E_2$, given in Fig.~\ref{table:jobelogic}(a), is functionally complete.
\end{Theorem}

\begin{figure}[!t]
  \subfigure[Over the set $\set{3,2,1}$]{
  \begin{minipage}{.4\columnwidth}
  \[
    \begin{array}{c|ccc|c|c}
        ~\bullet~ & ~3~ & ~2~ & ~1~ & ~E_1~ & ~E_2~ \\ \hline
        3     & 3 & 2 & 1 & 3   & 1 \\
        2     & 2 & 2 & 1 & 1   & 2 \\
        1     & 1 & 1 & 1 & 2   & 3
    \end{array}
  \]
  \end{minipage}
}
\hfill
  \subfigure[Over the set $\set{\match,\noreqmatch,\nomatch}$]{
  \begin{minipage}{.55\columnwidth}
  \[
    \begin{array}{c|ccc|c|c}
        ~\bullet~ & ~\match~ & ~\noreqmatch~ & ~\nomatch~ & ~E_1~ & ~E_2~ \\ \hline
        \match     & \match & \noreqmatch & \nomatch & \match   & \nomatch \\
        \noreqmatch     & \noreqmatch & \noreqmatch & \nomatch & \nomatch   & \noreqmatch \\
        \nomatch     & \nomatch & \nomatch & \nomatch & \noreqmatch   & \match
    \end{array}
  \]
  \end{minipage}
}
\caption{Jobe's 3-valued logic}\label{table:jobelogic}
\end{figure}

\newcommand{\swapop}{\mathop{\updownarrow}}

\begin{Cor}
The three-valued logic expressed over the set $\set{\nomatch,\match,\noreqmatch}$ and defined by the operators $\strongklor$, $\neg$ and $\kweakening$ is functionally complete.
\end{Cor}

\begin{proof}
We first define the operator $\strongkland$ from $\strongklor$ and $\lnot$: for any $X_1,X_2 \in \tdecisions$,
$(X_1 \strongkland X_2) = \lnot (\lnot X_1 \strongklor \lnot X_2)$\footnote{Note that we also have the expected equivalence $(X_1 \strongklor X_2) = \lnot (\lnot X_1 \strongkland \lnot X_2)$}.

We can clearly see from Fig.~\ref{table:jobelogic}(b), that the operator $\strongkland$ is identical to $\bullet$ and $\neg$ is identical to $E_1$.
Therefore, we only need to define a unary operator that swaps the values of $\nomatch$ and $\noreqmatch$ while leaving $\match$ unchanged.
We write $\swapop$ to denote such an operator.
The table below demonstrates that $\swapop X$ is equivalent to $(X \strongklor \noreqmatch) \strongkland (\kweakening (X \strongklor \lnot X))$.
  \[
    \begin{array}{c|c|c|c|c|c}
      X & X \strongklor \noreqmatch & \lnot X     & X \strongklor \lnot X & \kweakening (X \strongklor \lnot X)  & \swapop X \\ \hline
      \match      & \match      & \nomatch    & \match      & \match   & \match \\
      \nomatch    & \noreqmatch & \match      & \match      & \match   & \noreqmatch \\
      \noreqmatch & \noreqmatch & \noreqmatch & \noreqmatch & \nomatch & \nomatch \\
    \end{array}
  \]

We can therefore conclude that the logic defined over the set $\set{\nomatch,\match,\noreqmatch}$ by the operators $\strongklor, \lnot$ and $\kweakening$ is functionally complete.
\end{proof}

For instance, the operator $\tand$ can be built directly from $\tor$ and $\tnot$, since we can define the operator $\weakkland$ from $\strongklor$ and $\lnot$.
Indeed, for any $x,y \in \tdecisions$, we have the following equivalences:
\begin{align*}
    x \weakkland y & = (x \strongkland y) \strongklor ((x \strongkland \lnot x) \strongklor (y \strongkland \lnot y)) \\
    x \weakklor y & = (x \strongklor y) \strongkland ((x \strongklor \lnot x) \strongkland (y \strongklor \lnot y))
\end{align*}
We also have $x \kleenesup y = (x \strongklor (\kweakening y)) \strongkland ((\kweakening x) \strongklor y)$, where $\kleenesup$ is the supremum operator used to define the evaluation of an atomic target.
%We consider the functional completeness of our language in Section~\ref{}.

\section{Policies}\label{sec:policies}

PTaCL policies are defined inductively.
Let $d \in \set{\allow,\deny}$, and let $p$, $p_1$ and $p_2$ be policies. Then
\begin{itemize}
\item $d$ is a policy;
\item $\pnot p$ -- the \emph{negation} of policy $p$ -- is a policy, which returns $\allow$ if $p$ returns $\deny$ and vice versa;
\item $\dbd p$ -- the \emph{deny-by-default} of policy $p$ -- is a policy, which returns $\allow$ if $p$ returns $\allow$ and returns $\deny$ otherwise;
\item $p_1 \pand p_2$ -- the \emph{conjunction} of two policies $p_1$ and $p_2$ -- is a policy;
\item $(t,p)$ -- the \emph{restriction} of policy $p$ to a target $t$ -- is a policy.
\end{itemize}
We discuss policy evaluation in detail in Section~\ref{sec:policy-evaluation}.

A \emph{policy tree} is a convenient way of visualizing a policy and can be constructed recursively from a policy.
The policy $d$ is represented as a tree comprising a single node.
The policy $p_1 \pand p_2$ is represented as a tree comprising a root node labelled $\pand$ and two child sub-trees representing $p_1$ and $p_2$.
Policies of the form $(t,p)$, $\dbd p$ and $\pnot p$ are represented as trees comprising a root node labelled $t$, $\dbd$ and $\pnot$, respectively, a single child sub-tree representing $p$.
An illustrative policy tree representing the policy \[\dbd(t_5, \pnot(t_3,(t_1,\allow) \pand (t_2,\deny)) \pand (t_4,\allow))\] is shown in Fig.~\ref{fig:policytree}.
To save space, we have ``absorbed'' the nodes labelled $\pand$ into their respective parents ($t_3$ and $t_5$).

\subsection{Policy evaluation}\label{sec:policy-evaluation}

The evaluation of a policy with respect to a request $q$ returns $\na$ if the policy is not applicable to the request: that is, the evaluation of the policy's target with respect to $q$ returned $\nomatch$.
However, it may be the case that the evaluation of a target returns neither $\match$ nor $\nomatch$, instead returning $\noreqmatch$.
The possibility of target evaluation failing is considered in XACML~\cite{xacml2.0} and in the work of Li \emph{et al.}~\cite{li:acce09} and of Crampton and Huth~\cite{CH10}.
The methods used to handle such failures assume that target evaluation failures arise because of unexpected failures in hardware, software or network connectivity and, accordingly, make a best effort to construct a conclusive decision for the request.

Our target language is expressly designed to support flexible request formats for open environments.
As a result, our language explicitly includes the possibility that target evaluation may not be possible (if, for example, attributes are missing).
Hence, target evaluation may fail, not because of ``benign'' failures, but because a user may withhold attributes in an attempt to force an error in target evaluation and thereby circumvent policy evaluation.
Therefore, we must ensure that no advantage is gained by a malicious user who deliberately suppresses information when making an access request.%
\footnote{We also note the possibility that the user may not wish to divulge certain attributes when making an application request.}

Our approach is to consider all possible decisions that might have arisen had target evaluation not failed.
In other words, policy evaluation may return a set of decisions.
We shall see that imposing appropriate restrictions on targets and using a ``conservative'' method of deriving a single decision from a set of decisions, will enable us to guarantee that a malicious user obtains no advantage by withholding attribute information.

\newcommand{\pall}[1]{#1_{\sf all}}

% Given a policy $p$, we write $\pall{p}$ to denote the policy that is identical to $p$ except that its target has been replaced by $\tnull$.
% Then $p[\tnull]$ is identical to $p$ except that it is applicable to all targets.
% For brevity, we denote this policy by $\pall{p}$.
We recall the operators $\neg$, $\kweakening$ and $\strongkland$ on $\tdecisions$ (as shown in Fig.~\ref{fig:applicability-logical-operators}) and define the same operators on $\pdecisions = \set{\allow,\deny,\na}$.
We extend these unary operators to $X \subseteq \pdecisions$, writing $\neg X$ to denote the set $\set{\neg x : x \in X}$ and $\kweakening X$ to denote the set $\set{\kweakening x : x \in X}$; and we extend $\strongkland$ on $\pdecisions$ to sets $X,Y \subseteq \pdecisions$, writing $X \strongkland Y$ to denote the set $\set{x \strongkland y : x \in X, y \in Y}$.

Informally, the evaluation of targeted policy $(t, p)$ for a request $q$ proceeds in the following way.

% $p$ with request to a request $q$ proceeds in the following way.
%   \begin{compactenum}
%     \item If $p$ has the form $\pnot p'$ or $\dbd p'$, we evaluate $p'$ and then adjust the set of values returned as appropriate.
%     \item Otherwise, we evaluate whether the policy is applicable to $q$ or not:
\begin{compactenum}
\item If $t$ evaluates to $\match$,  we then inductively evaluate $p$ (see below)
\item If $t$ evaluates to $\nomatch$, we return $\set{\na}$
\item Otherwise, we evaluate $p$ and take the union of the resulting set of decisions with $\set{\na}$%
  \footnote{In other words, the evaluation of $p$ in this case considers the decisions that would have been returned if the request had been applicable and if the request had not been applicable.}
\end{compactenum}
% \end{compactenum}
We write $\psem{p}(q)$ to denote the evaluation of policy $p$ with respect to a request $q$, where
      \begin{align*}
        &\psem{d}(q) = d; \\
        &\psem{\pnot p}(q) = \neg (\psem{p}(q)); \\
        &\psem{\dbd p}(q) = \kweakening (\psem{p}(q)) \\
    &\psem{(p_1 \pand p_2)}(q) = \psem{p_1}(q) \strongkland \psem{p_2}(q);\\
    &\psem{(t, p)}(q) =
      \begin{cases}
        \psem{p}(q) & \text{if $\tsem{t}(q) = \match$}, \\
        \set{\na} & \text{if $\tsem{t}(q) = \nomatch$}, \\
        \set{\na} \cup \psem{p}(q) & \text{otherwise}.
      \end{cases}
  \end{align*}

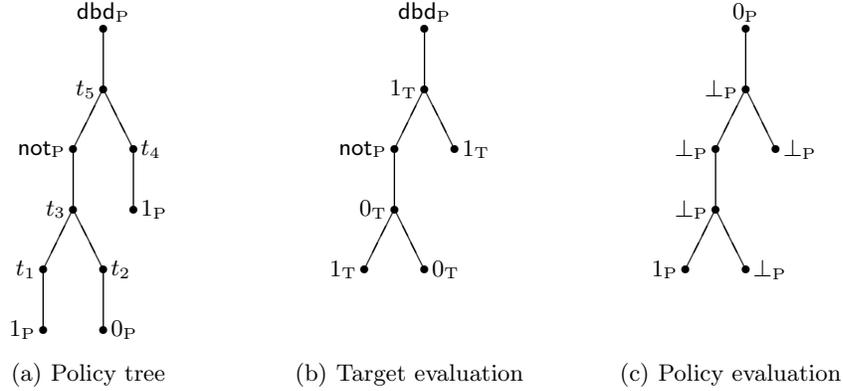
\begin{figure}[h]
\subfigure[Policy tree]{\label{fig:policytree}%
\begin{minipage}{.3\textwidth}\centering\setlength{\unitlength}{.2mm}
  \begin{picture}(60,220)(-40,-50)
    \node{r}{-40}{-40}{$\allow$}
      \drawline(-40,-40)(-40,0)
    \node{l}{0}{-40}{$\deny$}
      \drawline(0,-40)(0,0)
    \node{r}{-40}{0}{$t_1$}
      \drawline(-40,0)(-20,40)
    \node{l}{0}{0}{$t_2$}
      \drawline(0,0)(-20,40)
    \node{r}{-20}{40}{$t_3$}
      \drawline(-20,40)(-20,80)
    \node{r}{-20}{80}{$\pnot$}
      \drawline(-20,80)(0,120)
    \node{l}{20}{40}{$\allow$}
      \drawline(20,40)(20,80)
    \node{l}{20}{80}{$t_4$}
      \drawline(20,80)(0,120)
    \node{r}{0}{120}{$t_5$}
      \drawline(0,120)(0,160)
    \node{b}{0}{160}{$\dbd$}
  \end{picture}
\end{minipage}}
\hfill
\subfigure[Target evaluation]{\label{fig:targeteval}%
\begin{minipage}{.3\textwidth}\centering\setlength{\unitlength}{.2mm}
  \begin{picture}(60,220)(-40,-50)
    \node{r}{-40}{0}{$\match$}
      \drawline(-40,0)(-20,40)
    \node{l}{0}{0}{$\nomatch$}
      \drawline(0,0)(-20,40)
    \node{r}{-20}{40}{$\nomatch$}
      \drawline(-20,40)(-20,80)
    \node{r}{-20}{80}{$\pnot$}
      \drawline(-20,80)(0,120)
    \node{l}{20}{80}{$\match$}
      \drawline(20,80)(0,120)
    \node{r}{0}{120}{$\match$}
      \drawline(0,120)(0,160)
    \node{b}{0}{160}{$\dbd$}
  \end{picture}
\end{minipage}}
\hfill
\subfigure[Policy evaluation]{\label{fig:policyeval}%
\begin{minipage}{.3\textwidth}\centering\setlength{\unitlength}{.2mm}
  \begin{picture}(60,220)(-40,-50)
    \node{r}{-40}{0}{$\allow$}
      \drawline(-40,0)(-20,40)
    \node{l}{0}{0}{$\na$}
      \drawline(0,0)(-20,40)
    \node{r}{-20}{40}{$\na$}
      \drawline(-20,40)(-20,80)
    \node{r}{-20}{80}{$\na$}
      \drawline(-20,80)(0,120)
    \node{l}{20}{80}{$\na$}
      \drawline(20,80)(0,120)
    \node{r}{0}{120}{$\na$}
      \drawline(0,120)(0,160)
    \node{b}{0}{160}{$\deny$}
  \end{picture}
\end{minipage}}
\caption{Evaluating a PTaCL policy}
\end{figure}

Consider the policy depicted in Fig.~\ref{fig:policytree} and suppose that $\sem{t_1}(q) = \sem{t_4}(q) = \sem{t_5}(q) = \match$, $\sem{t_2}(q) = \nomatch$ and $\sem{t_3} = \noreqmatch$.
The evaluation of this policy is shown in Fig.~\ref{fig:policyeval}.
Note that the evaluation of the sub-tree with root $t_3$ considers the union of two sets of decisions because $\sem{t_3}(q) = \noreqmatch$.
Note also that the strong conjunction $\strongkland$ has the effect of preferring the $\na$ decision to the $\allow$ decision.
For those familiar with previous related work, this may seem an unusual way in which to combine policy decisions.
We discuss this in more detail in the next section and, in Section~\ref{sec:decisionops}, we will discuss ways in which more familiar decision-combining operators can be defined.
Finally, note that the policy does evaluate to a single decision ($\deny$) for this request, although there is no reason in general for this to occur.
However, it is easy to establish the following result.

\begin{Lem}
Let $p$ be a policy whose policy tree contains targets $t_1,\dots,t_k$ and let $q$ be a request.
If $\sem{t_i}(q) \ne \noreqmatch$ for all $i$, then $\sem{p}(q) = \set{x}$ for some $x \in \pdecisions$.
\end{Lem}

In other words, if the applicability of all targets referenced by a policy can be determined for a request $q$, our evaluation semantics will return a unique authorization decision.
The proof is a straightforward induction on the depth of the policy tree.

Finally, we note that the functional completeness for the target language also holds for our policy language, because $\tweakening_{\rm T}$ and $\dbd$ have identical properties, as do $\tnot_{\rm T}$ and $\pnot$.
However, it is also important to realize that the interpretation of $\noreqmatch$ and $\na$ are quite different: the former indicates that the request supplied insufficient information to evaluate target applicability, whereas $\na$ indicates that a policy is irrelevant to the evaluation of a request.
Henceforth, we will omit the subscript from $\psem{\cdot}$ and the PTL operators, although, for clarity, we will retain the subscripts on decisions.

\renewcommand{\pand}{\mathbin{\mathsf{and}}}
\renewcommand{\dbd}{\mathop{\mathsf{dbd}}}
\renewcommand{\pnot}{\mathop{\mathsf{not}}}

\subsection{On the non-monotonicity of targets}

The language we use for targets and the way in which targets are evaluated means that, for some target $t$, there may exist requests $q$ and $q'$ such that $q' \subseteq q$, $\sem{t}(q') = \nomatch$ and $\sem{t}(q) = \match$.
This feature of the language means that withholding attributes may provide some advantage to a malicious user: if we have a policy $p = (t,p')$ such that $\sem{p'}(q) = \deny$, and $\sem{t}(q) = \match$, then $\sem{p}(q) = \deny$; if, however, $\sem{t}(q) = \nomatch$, then $\sem{p}(q) = \na$.
In other words, it might be possible for a malicious user to turn a $\deny$ decision into a $\na$ decision by suppressing certain attributes.
For brevity, we refer to this as the non-monotonicity of targets.
Hence, we might reasonably regard $\na$ as a potentially dangerous policy decision.
(This view of $\na$ is quite different from the interpretation used by other policy languages and algebras.)
It is this view that informs our use of $\strongkland$ to combine policy decisions, which means that $\na \strongkland \allow$ is defined to be $\na$ rather than $\allow$.

Similarly, a user can force a target to evaluate to $\noreqmatch$ (rather than $\nomatch$ or $\match$) by withholding attributes.
It is for this reason, that policy evaluation considers the possibility that a target might have been matched or not matched when target evaluation returns $\noreqmatch$.

Following from the above discussion, we would like to prove a result of the form:
{\em Let $p$ be a policy whose policy tree contains targets $t_1,\dots,t_k$ and let $q$ be a request.
Then for any $q' \subseteq q$, $\sem{p}(q) \subseteq \sem{p}(q')$.}
Informally, this result states that if a request contains less information, then the result of evaluating the policy is more uncertain.
Then the authorization decision point can have a decision-set ``resolution strategy'' that returns a single final decision.
Such a strategy should be ``conservative'' in the sense that the larger decision sets should be treated with more caution.
The obvious strategy of this nature is: for $X \subseteq \pdecisions$, we return $\allow$ if $X = \set{\allow}$ and $\deny$ otherwise.
%An alternative strategy is to try to recover more attributes, thereby reducing the uncertainty in the set of decisions returned by policy evaluation.

However, it is easy to see that the above result does not hold, because of the functional completeness of our target language.
In particular, we can create an operator $\oplus$ such that $\noreqmatch \oplus \noreqmatch = \match$ and $\match \oplus \noreqmatch = \noreqmatch$.
Now consider the target $t = (n_1, v_1) \oplus (n_2, v_2)$, and the requests $q_1 = \set{(n_1, v_1)}$ and $q_2 = \set{}$.
Then
  \[
    \sem{t}(q_1) = \match \oplus \noreqmatch = \noreqmatch \quad\mbox{and}\quad \sem{t}(q_2) = \noreqmatch \oplus \noreqmatch = \match.
  \]
Now consider the policy $p = (t, \allow)$: we have $\sem{p}(q_1) = \set{\na, \allow}$ and $\sem{p}(q_2) = \set{\allow}$, providing a counter-example to the desired result.
In other words, there are good reasons to restrict our target language so that only ``well-behaved'' targets can be defined.
Specifically, we would like to restrict our target language so that all targets have the following property:

\begin{Def}
A target $t$ is \emph{monotonic} if for all requests $q$ and for every $q' \subseteq q$, $\sem{t}(q') \in \set{\noreqmatch, \sem{t}(q)}$.
\end{Def}
Then we have the following result (the proof is given in Appendix~\ref{app:monotonicity} and has been encoded in 
the proof assistant Isabelle/Isar\footnote{\url{http://isg.rhul.ac.uk/~jason/isabelle/ptacl.thy}}).

\begin{Thm}\label{thm:strong-monotonicity}
Let $p$ be a policy whose policy tree contains \emph{monotonic} targets $t_1,\dots,t_k$ and let $q$ be a request.
Then for any $q' \subseteq q$, $\sem{p}(q) \subseteq \sem{p}(q')$.
\end{Thm}

The obvious questions to ask now are: Which of our target operators are monotonic? And does composition of monotonic target operators preserve monotonicity?

We say that an operator is monotonic if, given monotonic targets as inputs, it returns a monotonic target. We prove in appendix~\ref{app:monotonicity} that the operators
$\tnot, \tand$ and $\tor$ are monotonic, as well as the operators corresponding to $\strongkland$ and $\weakklor$.
However, the operator $\tweakening$ is not monotonic, since it can transform a $\noreqmatch$ into a $\nomatch$.

Unfortunately (and somewhat unexpectedly), an atomic target is not, in general, monotonic.
To see this, note that a request can contain several pairs with the same attribute name.
(A request might, for example, enumerate all the roles with which the requester is associated.)
Removing one occurrence from this set of pairs can change the evaluation of the request from $\match$ to $\nomatch$.
This situation corresponds to a partial hiding of attribute values: that is, the ability for a user or an attribute server to remove only some values for a given attribute.
In practice, such a situation is quite hard to detect and to prevent.
However, let us assume that an attribute server works in an ``all-or-nothing mode'': that is, either it returns all the values for a given attribute, or none.
With this assumption, for two requests $q$ and $q'$ such that $q' \subseteq q$ and for any attribute name $n$ such that $(n, v) \in q'$ and $(n, v') \in q$, then $(n, v') \in q'$.
With such an assumption, it is easy to see that any atomic target is monotonic, and it follows that any target built using the operators $\tand$, $\tor$ and $\tnot$ is monotonic.

Such an assumption might not always hold, in particular when there is little control over the attribute servers.
Therefore, we now consider an alternative, weaker notion of monotonicity, defined below.

\begin{Def}
A target $t$ is \emph{weakly monotonic} if for all requests $q$ and for every $q' \subseteq q$,  $\sem{t}(q') \preccurlyeq \sem{t}(q)$, where we define $\noreqmatch \prec \nomatch \prec \match$.
\end{Def}

The operators $\kweakening, \weakkland, \weakklor$ and $\strongklor$ preserve the weak monotonicity, as proven in Appendix~\ref{app:weak-monotonicity}, but the operators $\lnot$ and $\strongkland$ do not.
Moreover, since any atomic target is clearly weakly monotonic, any target built using any combination from the operators $\kweakening, \weakkland, \weakklor$ and $\strongklor$ is also weakly monotonic.
Although we cannot prove a result as strong as Theorem~\ref{thm:strong-monotonicity}, we can prove the following result (the proof of which can be found in Appendix~\ref{app:weak-monotonicity}).

\begin{Thm}\label{thm:weak-monotonicity}
Let $p$ be a policy whose policy tree contains \emph{weakly monotonic} targets $t_1,\dots,t_k$ and let $q$ be a request.
  \begin{compactenum}
    \item If $p$ is constructed from the operators $\pnot$ and $\pand$, then for any $q' \subseteq q$, if $\sem{p}(q') = \set{d}$, with $d \in \set{\allow, \deny}$, then $\sem{p}(q) = \sem{p}(q')$.
    \item If $p$ is constructed from the operators $\dbd$ and $\pand$, then for any $q' \subseteq q$, if $\sem{p}(q') = \set{\allow}$, then \mbox{$\sem{p}(q) =  \set{\allow}$}.
  \end{compactenum}
\end{Thm}

One consequence of Theorem~\ref{thm:weak-monotonicity} is that if a partial request is allowed, then the full request would have been allowed too, and therefore an attacker has no advantage in hiding some attribute values.
However, this result requires a ``conservative'' resolution strategy: that is, request $q$ is only allowed if and only if $\sem{p}(q) = \set{\allow}$.
%, and that the policy is defined without the operator $\dbd$, and that the targets of the policy are defined using operators in $\set{\kweakening, \weakkland, \weakklor, \strongklor}$.

\subsection{Decision operators}\label{sec:decisionops}

We now discuss other ways in which decisions from sub-policies might be combined.
Following Crampton and Huth~\cite{CH10}, we restrict attention to \emph{idempotent} and \emph{well-behaved} decision operators.
%We introduce these operators for the decision set $\set{\allow,\deny,\na}$; we extend their definition to $\set{\allow,\deny,\na,\undef}$ in the next section.

\begin{Def}
Let $\oplus : \pdecisions \times \pdecisions \rightarrow \pdecisions$ be a decision operator.
  \begin{compactitem}
    \item If $x \oplus x = x$ for all $x \in \pdecisions$, then we say $\oplus$ is \emph{idempotent}.
    \item If $x \oplus \na = x = \na \oplus x$ for all $x \in \pdecisions$, then we say $\oplus$ is a \emph{$\cup$-operator}.
    \item If $x \oplus \na = \na = \na \oplus x$ for all $x \in \pdecisions$, then we say $\oplus$ is an \emph{$\cap$-operator}.
    \item We say $\oplus$ is \emph{well-behaved} if it is either a $\cup$- or an $\cap$-operator.
  \end{compactitem}
\end{Def}

Informally, a $\cup$-operator ignores policies that evaluate to $\na$ by returning a conclusive decision (that is, a decision that belongs to $\set{\allow,\deny}$) if either operand returns a conclusive decision.
XACML, for example, assumes that all operators are $\cup$-operators.
In contrast, a $\cap$-operator only returns a conclusive decision if both arguments are conclusive decisions.
An operator of this nature is used by Bonatti \etal in their policy algebra~\cite{bona:alge02}.

Intuitively, it seems reasonable to assume that a policy decision operator is idempotent: if two policies return the same decision $d$, then we would expect that the composition of those policies would also return $d$.
An idempotent, well-behaved decision operator is uniquely defined by the choices of $x \oplus \na$, $\allow \oplus \deny$ and $\deny \oplus \allow$: the remaining values are fixed because the operator is idempotent and well-behaved (as shown in Fig.~\ref{fig:decisionoperators} for an idempotent $\cup$-operator $\oplus$).

If we assume that $\oplus$ is commutative, then there are only three choices for an idempotent $\cup$-operator (and three choices for an idempotent $\cap$-operator).
And if we assume that $\allow \oplus \deny \in \set{\allow,\deny}$, then there are only two choices for a commutative, idempotent $\cup$-operator; both these operators are shown in Fig.~\ref{fig:decisionoperators}, labeled as $\andcup$ and $\orcup$.
Analogous operators $\andcap$ and $\orcap$ can be defined by making the obvious adjustments to the bottom row and rightmost column of the tables for $\andcup$ and $\orcup$, respectively.

The operators $\andcup$ and $\andcap$ are rather similar to logical conjunction, while $\orcup$ and $\orcap$ are rather similar to logical disjunction, respectively.
Our decision operators play a similar role to the conflict resolution strategies or policy-combining algorithms used in policy algebras and XACML.
Such strategies are used to resolve discrepancies in the results returned by different sub-policies.
In particular, $\andcup$ has the same effect as the ``deny-overrides'' conflict resolution strategy: namely, if one sub-policy returns $\deny$, then the combined decision is $\deny$.
Similarly, $\orcup$ has the same effect as the ``allow-overrides'' strategy.
%
%\begin{Pro}
%If $\oplus_1$ and $\oplus_2$ are commutative, idempotent and well-behaved, then the composition of $\oplus_1$ and $\oplus_2$ is commutative, idempotent and well-behaved.
%\end{Pro}

%If $\oplus$ is idempotent and well-behaved but not commutative, then there are six $\cup$-operators and six $\cap$-operators; there are two choices for each type if $\allow \oplus \deny \in \set{\allow,\deny}$ and $\deny \oplus \allow \in \set{\allow,\deny}$.
The most widely used non-commutative conflict resolution strategy is ``first-applicable'', which we denote by $\rhd$.
The operator $\rhd$ is defined in Fig.~\ref{fig:decisionoperators}(d): note, in particular, $\allow \rhd \deny = \allow$ and $\deny \rhd \allow = \deny$.%
\footnote{Note that a first-applicable $\cap$-operator is vacuous, as it would be equivalent to a unary, identity operator.}
The first-applicable operator is commonly used in firewall rulesets as well as in policy algebras and XACML.
The other idempotent, well-behaved, non-commutative operator such that $\allow \oplus \deny \in \set{\allow,\deny}$ and $\deny \oplus \allow \in \set{\allow,\deny}$ is what might be called ``last-applicable'', denoted by $\lhd$, where $x \lhd y = y$ if $y \in \set{\allow,\deny}$ and is equal to $x$ otherwise.
This operator does not appear to be widely supported or used.

\begin{figure*}[!t]
    \subfigure[Idempotent]{
    \begin{minipage}{.23\textwidth}
  \[
    \begin{array}{c|ccc}
      \oplus & \allow & \deny & \na \\
    \hline
      \allow & \allow & x & \allow \\
      \deny & y & \deny & \deny \\
      \na & \allow & \deny & \na
    \end{array}
    \]
    \end{minipage}}
    \hfill
  \subfigure[Conjunction]{
  \begin{minipage}{.23\textwidth}
    \[
    \begin{array}{c|ccc}
      \andcup & \allow & \deny & \na \\
    \hline
      \allow & \allow & \deny & \allow \\
      \deny & \deny & \deny & \deny \\
      \na & \allow & \deny & \na
    \end{array}
    \]
    \end{minipage}}
    \hfill
    \subfigure[Disjunction]{
      \begin{minipage}{.23\textwidth}
    \[
    \begin{array}{c|ccc}
      \orcup & \allow & \deny & \na \\
    \hline
      \allow & \allow & \allow & \allow \\
      \deny & \allow & \deny & \deny \\
      \na & \allow & \deny & \na
    \end{array}
    \]
    \end{minipage}}
    \hfill
  \subfigure[First-applicable]{
      \begin{minipage}{.23\textwidth}
    \[
    \begin{array}{c|ccc}
      \rhd & \allow & \deny & \na \\
    \hline
      \allow & \allow & \allow & \allow \\
      \deny & \deny & \deny & \deny \\
      \na & \allow & \deny & \na
    \end{array}
    \]
    \end{minipage}}
\caption{Decision tables for idempotent $\cup$-operators on $\set{\allow,\deny,\na}$}
\label{fig:decisionoperators}
\end{figure*}

We now show how to define the operators $\orcap$, $\andcap$, $\orcup$, $\andcup$ and $\rhd$ from the PTL operators $\pnot$, $\dbd$ and $\pand$.
Since the logic $(\set{\allow, \deny, \na}, \pnot, \dbd, \pand)$ is functionally complete, we can directly reuse the definitions of the operators given in Fig.~\ref{fig:applicability-logical-operators}.
Clearly, $\orcap$ and $\andcap$ are directly given by $\weakklor$ and $\weakkland$, respectively.
Moreover, the operator $\orcup$ corresponds to the supremum operator over the total order $\allow > \deny > \na$, so we can re-use the operator $\kleenesup$ defined in Section~\ref{sec:targetcompleteness}. The operator $\andcup$ is defined as the double negation of the operator $\orcup$:
\[
  x \andcup y = \pnot((\pnot x) \orcup (\pnot y))
\]
In order to define the operator $\rhd$, we first introduce the operator $\abd$ (``allow-by-default''), which transforms  $\na$ into $\allow$, and is defined by $\abd x = \pnot (\dbd (\pnot x))$.
The definition of $\rhd$ is then given by:
\[
  x \rhd y = (\abd (x \strongklor (\pnot x))) \strongkland (x \orcup y)
\]
Finally, $x \lhd y$ is equivalent to $y \rhd x$.
Henceforth, we will use the operators defined above as syntactic sugar.
Notice that our definitions of $\orcup$, $\andcup$ and $\rhd$ all require the three PTL operators for their construction.
Hence, a policy containing the standard XACML operators does not satisfy the requirements of Theorem~\ref{thm:weak-monotonicity}, so we need to rely on the all-or-nothing assumption.

Finally, we note that the operators $\pand$, $\andcup$ and $\andcap$ can be regarded as defining a greatest lower bound operator for suitable choices of ordering on $\pdecisions$; similarly $\orcup$ and $\orcap$ define least upper bound operators.
These orderings are summarized in Table~\ref{tbl:decision-operators-as-glbs-lubs}.

\begin{table}[h]
  \[
    \begin{array}{|c|c|}
    \hline
      \mbox{Operator} & \mbox{Ordering} \\
    \hline
      \pand & \deny < \na < \allow \\
      \andcup & \deny < \allow < \na\\
      \andcap & \na < \deny < \allow \\
      \orcup & \na < \deny < \allow \\
      \orcap & \deny < \allow < \na   \\
    \hline
    \end{array}
  \]
\caption{Decision operators and orderings on $\pdecisions$}\label{tbl:decision-operators-as-glbs-lubs}
\end{table}

The fact that each of the orderings is a total order means that $\pand$, $\andcup$ and $\andcap$ take the minimum of their operands, while $\orcup$ and $\orcap$ take the maximum of their operands.
This, in turn, means that all four operators can be extended to $n$-ary operators (for any natural number $n > 1$).
We examine the consequences of this in the next section.

\subsection{Policy sets and policy lists}

\begin{figure}[h]  \setlength{\unitlength}{.8pt}\centering
\subfigure[Binary operators]{\label{subfig:binarypolicytree}
  \begin{picture}(120,100)(-60,-10)
    \node{b}{0}{80}{$(t,\orcup)$}
      \drawline(0,80)(-40,40) \drawline(0,80)(40,40)
    \node{r}{-40}{40}{$(t,\orcup)$}
      \drawline(-40,40)(-60,0) \drawline(-40,40)(-20,0)
    \node{r}{40}{40}{$(t_3,\allow)$}
    \node{t}{-60}{0}{$(t_1,\allow)$}
    \node{t}{-20}{0}{$(t_2,\deny)$}
  \end{picture}}
  \qquad
\subfigure[$n$-ary operators]{\label{subfig:narypolicytree}
  \begin{picture}(100,55)(-50,-10)
    \node{b}{0}{40}{$(t,\orcup)$}
      \drawline(0,40)(-50,0)
      \drawline(0,40)(0,0)
      \drawline(0,40)(50,0)
    \node{t}{-50}{0}{$(t_1,\allow)$}
    \node{t}{0}{0}{$(t_2,\deny)$}
    \node{t}{50}{0}{$(t_3,\allow)$}
  \end{picture}}
\caption{From binary to $n$-ary operators}\label{fig:combining-policies-with-commutative-ops}
\end{figure}
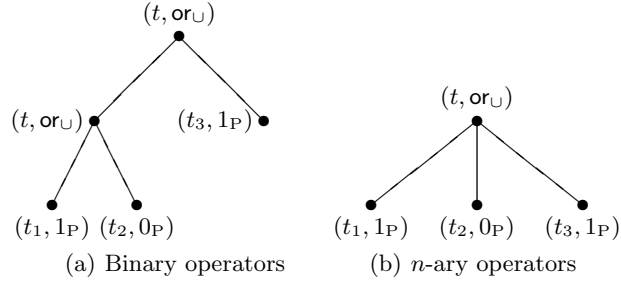

Consider the policy tree depicted in Figure~\ref{subfig:binarypolicytree}.
This policy can be expressed as
  \[
    \Big(t,\big(t,(t_1,\allow) \orcup (t_2,\deny)\big) \orcup (t_3,\allow)\Big)
    %\Big(t,\orcup,\big(t,\orcup,(t_1,\allow),(t_2,\deny)\big),(t_3,\allow)\Big).
  \]
Given that the target and operator match in the two non-leaf nodes and that $\orcup$ is commutative, we could represent the policy as
  \[
    \big(t,\orcup,\set{(t_1,\allow),(t_2,\deny),(t_3,\allow)}\big),
  \]
as illustrated in Figure~\ref{subfig:narypolicytree}.

Clearly, we can extend this argument to policy trees that combine $n$ subtrees with equivalent targets and use the same commutative operator.
Accordingly, we extend our policy syntax to include policies of the form $(t,\oplus,P)$, where $t$ is a target, $\oplus$ is a commutative operator and $P$ is a set of policies.
Then, for all $k \geqslant 2$, for all policies $p_1,\dots,p_k$ and all operators $\oplus \in \set{\pand,\andcup,\orcup,\andcap,\orcap}$ we define
  \[
    \sem{(\tnull,\oplus,\set{p_1,\dots,p_k})}(q) = \set{d_1 \oplus \dots \oplus d_k : d_i \in \sem{p_i}(q)},
  \]
Now we recall that $\pand$, $\andcup$ and $\andcap$ can be interpreted as greatest lower bound operators defined over suitable total orderings on $\set{\allow,\deny,\na}$ (as shown in Table~\ref{tbl:decision-operators-as-glbs-lubs}).
Similarly, $\orcup$ and $\orcap$ are least upper bound operators.
Hence, the evaluation of $d_1 \oplus \dots \oplus d_k$ is equivalent to computing the maximum value in the set $\set{d_1,\dots,d_k}$ in the case of disjunctive operators and the minimum value in the case of conjunctive operators.
Consider, for example,
  \[
    \set{d_1 \andcup \cdots \andcup d_k : d_i \in \sem{p_i}(q)}.
  \]
Now we know that $x_1 \andcup \cdots \andcup x_k$ is the minimum value in $\set{x_1,\dots,x_k}$ with respect to the order $\deny < \allow < \na$.
Hence, the maximum value that $d_1 \andcup \cdots \andcup d_k$ can take is $d_{\max}$, where
  \[
    d_{\max} = \min\set{\max(\sem{p_1}(q)),\dots,\max(\sem{p_k}(q))}.
  \]
Similarly, the minimum value that $d_1 \andcup \cdots \andcup d_k$ can take is $d_{\min}$, where
  \[
    d_{\min} = \min\set{\min(\sem{p_1}(q)),\dots,\min(\sem{p_k}(q))}.
  \]
Now if $d_{\min} = d_{\max}$, we have
  \[
    \set{d_1 \andcup \cdots \andcup d_k : d_i \in \sem{p_i}(q)} = \set{d_{\min}};
  \]
otherwise, $\set{d_1 \andcup \cdots \andcup d_k : d_i \in \sem{p_i}(q)}$ is a set containing more than one decision and the conservative approach requires us to return $\deny$.
If, for example, $\sem{p_1}(q) = \set{\deny,\allow}$, $\sem{p_2}(q) = \set{\allow}$ and $\sem{p_3}(q) = \set{\deny,\na}$, then $d_{\max} = \allow$ and $d_{\min} = \deny$.

We can derive analogous methods for computing $\set{d_1 \oplus \cdots \oplus d_k : d_i \in \sem{p_i}(q)}$, for $\oplus \in \set{\andcap,\orcup,\orcap}$.
Hence, we can compute $\set{d_1 \oplus \cdots \oplus d_k : d_i \in \sem{p_i}(q)}$ in linear time for $\oplus \in \set{\pand,\andcup,\andcap,\orcup,\orcap}$.

Note, however, that we cannot apply the same re-writing technique to the policy
  \[
    \Big(t,\big(t,(t_1,\allow) \rhd (t_2,\deny)\big) \rhd (t_3,\allow)\Big)
  \]
because $\rhd$ is not commutative (so the order in which sub-policies are evaluated is significant).
Hence, we need to interpret the policy as
  \[
    (t,\rhd,[(t_1,\allow),(t_2,\deny),(t_3,\allow)]),
  \]
where $[p_1,\dots,p_k]$ is to be understood as a list of sub-policies that must be evaluated in the specified order.
Moreover, because $\rhd$ has no interpretation as a maximum or minimum operator, we have no quick way of evaluating
  \[
    \set{d_1 \rhd \cdots \rhd d_k : d_i \in \sem{p_i}(q)}
  \]
because we have to evaluate each possible combination of $d_1,\dots,d_k$.
To summarize:
  \begin{compactitem}
    \item We can group a set or list of policies together under a single target and decision operator.
    %\item The evaluation of a policy may terminate early if a decision is found that renders subsequent processing irrelevant: if, for example, we find that a sub-policy evaluates to $\allow$ and the operator is $\orcup$ then we can return $\allow$; or if we find that a sub-policy evaluates to $\na$ and the operator is $\orcap$ then we can return $\na$.
    \item All commutative operators can use either set or list processing; non-commutative operators can only use list processing.
    \item Policy sets can be evaluated in any order, whereas policy lists must be processed in the order specified in the list.
    \item Most importantly, policy sets can be evaluated quickly, unlike policy lists (for non-commutative operators).
  \end{compactitem}

\subsection{Policy equivalence}

We say two policies $p_1$ and $p_2$ are equivalent if for all requests $q$, $\sem{p_1}(q) = \sem{p_2}(q)$.
We write $\sem{p_1} = \sem{p_2}$ if $p_1$ and $p_2$ are equivalent.

Note that $(t,\allow)$ returns $\allow$ for all requests that match $t$, so $\pnot (t,\allow)$ returns false for all requests that match $t$.
In other words, the policy $(t,\deny)$ is equivalent to the policy $\pnot (t,\allow)$, which means that the policies of the form $(t,\deny)$ are not required (although they may be useful as syntactic sugar).

\begin{Pro}
For all policies $p_1$ and $p_2$,
  \begin{align*}
    \sem{\dbd(p_1 \pand p_2)} &= \sem{(\dbd p_1) \pand (\dbd p_2)}, \\
    \sem{\pnot(p_1 \andcup p_2)} &= \sem{(\pnot p_1) \orcup (\pnot p_2)}, \\
    \sem{\pnot(p_1 \orcup p_2} &= \sem{(\pnot p_1) \andcup (\pnot p_2)}, \\
    \sem{\pnot(p_1 \andcap p_2)} &= \sem{(\pnot p_1) \orcap (\pnot p_2)}, \\
    \sem{\pnot(p_1 \orcap p_2} &= \sem{(\pnot p_1) \andcap (\pnot p_2)}.
  \end{align*}
\end{Pro}

\begin{proof}
The proofs follow by inspection of the appropriate decision tables.
\end{proof}

\section{Related work}

It is important to note that \ptackle is neither intended to fix XACML nor to provide formal semantics for XACML policy evaluation.
Rather, \ptackle is a language that seeks to provide rigorous, alternative solutions to the same problems that motivated the development of XACML.
Our work is also influenced by the work of Li {\em et al.}~\cite{li:acce09} and of Crampton and Huth~\cite{CH10} on using a set of decisions, rather than a single decision, to define the result of policy evaluation.

Although there is a substantial body of work on policy specification~\cite{back:alge04,bona:alge02,BH08,ni:dalg09,wije:prop03}, this prior work assumes a very restricted format for access requests and targets.
To the best of our knowledge, there is no previous work on a formal language for target specification and evaluation, let alone the consideration of missing attributes names.
Both the ratified standard XACML 2.0~\cite{xacml2.0} and the draft XACML 3.0~\cite{XACML3}, acknowledge that attributes may be missing from a request.
However, the treatment of target evaluation in such circumstances is, like much of the XACML standard, rather informal.
Moreover, the XACML target syntax is unnecessarily complicated and does not support interface targets.
Finally, the XACML target syntax only provides operators that are equivalent to the strong conjunction and strong disjunction (in the $3$-value Kleene logic), thereby limiting the expressive power of XACML.
On the other hand, the functional completeness of PTL means that any XACML target can be represented in PTL.

The work on policy algebras varies in the operators that are supported, the set of decisions that can arise as a result of policy evaluation, and the extent to which policy evaluation can cope with failures in target evaluation.
Ni {\em et al.}, for example, provide a functional complete policy algebra~\cite{ni:dalg09}, where policy evaluation returns a single decision from the set $\set{\allow,\deny,\na}$.
The functional completeness of PCL means that we can express any operators that we might wish to.
In particular, we can express all XACML policy-combining algorithms.
Structurally, our atomic policies correspond to rules in XACML, while our policy trees correspond to policies and policy sets.
Crampton and Huth~\cite{CH10} extend the work of Li {\em et al.} on policy evaluation in the presence of target evaluation failure~\cite{li:acce09}, where policy evaluation returns a set of decisions.
Our treatment of policy evaluation is rather similar to this earlier work, although the way in which we resolve a set of decisions to a single decision that is enforced by the AEF is completely different, due to the suspicion with which we choose to treat the $\na$ decision.

An important contribution of this paper is the recognition that providing support for attribute-based access control and greater freedom for request formats leads to the potential for attribute hiding by malicious users.
By manipulating requests in this way, it may be possible to circumvent the expected or intended policy semantics.
Existing work that supports attribute-based access control, such as XACML 3.0 and that of Rao {\em et al.}~\cite{Rao:2009:AFI:1542207.1542218}, does not consider such possibilities and hence may be vulnerable to ``attribute-hiding attacks''.
Consider, for example, the PTL policy $p = (\allow \andcup ((n, v), \deny))$~--~which corresponds to an XACML policy with two rules combined using the deny-overrides operator~--~and two requests $q = \set{(n, v), (n, v')}$ and $q'=\set{(n, v')}$.
Then $\sem{p}(q) = \deny$ while $\sem{p}(q') = \allow$: that is, by hiding some information, a more favorable answer is obtained.
Theorem~\ref{thm:weak-monotonicity} suggests that such behavior is to be expected because we require all three PTL operators to represent $\andcup$.

\section{Concluding remarks}

Attribute-based access control, rather than the traditional identity-based access control that is deployed extensively in closed systems, is likely to become increasingly important in loosely coupled and open computing environments.
This paper introduces \ptackle, an expressive language for the definition of attribute-based authorization policies.
\ptackle can represent all commonly used policy composition operators (indeed it can represent any desired operator) and, to the best of our knowledge, \ptackle is the first language with a concise syntax for policy targets and a precise semantics for their evaluation.
%Moreover, we provide formal evaluation functions, based on well-known logical properties, allowing us, for instance, to evaluate sets of policies efficiently.
%Due to the lack of space, we do not consider here obligations, but they can be easily integrated in \ptackle by a straight-forward extension of the policy syntax.

Nevertheless, \ptackle is rather simple syntactically, which enables us to identify and propose solutions to the problem of attribute hiding.
% an issue related to the hiding, partial or complete, of information from the context, potentially leading a partial request to be granted while the full request is not.
Such an issue is problematic in the context of open and distributed systems, and is not addressed in the literature, which define composition operators to favor conclusive decisions over a not-applicable decision.
Having identified the problem, we propose two approaches to address this issue, formally justifying each of them: either forbidding optional targets, assuming the attribute servers to work in an ``all-or-nothing mode'' and adopting a conservative evaluation; or constraining more strictly the definition of the targets and the definition of the policies.
The second approach does not make any assumption of the attribute servers, but the standard policy composition operators can no longer be used.
We propose other operators that are resilient to attribute hiding and differ from the standard ones in the way in which they handle the not-applicable decision.
These ``new'' operators actually correspond to the strong conjunction and strong disjunction defined in the original Kleene three-valued logic.

There are many opportunities for future work. Clearly, when 
the evaluation of a request returns more than one decision, it 
implies that some attributes are missing in the request, and \ptackle should be extended in order for the set of the decisions to also indicate which attributes are missing. Hence, the entity in charge of collecting the attributes, for instance the Context Handler in the XACML architecture, is able know which attributes to collect again. PCL can be similarly extended in order to support obligations, that can be returned in addition to a set of decisions. 

These extensions naturally lead to the problem of understanding and formalizing the complete access control architecture, and in particular to the question of {\em attribute privacy}. Indeed, in practice, a reason for a missing attribute can be because the source responsible for providing its value considered that this value was too sensitive to be shared. In such a case, the evaluation of the policy, or part of it, needs to be delegated to the attribute source. However, the possible presence of multiple, sensitive and conflicting sources makes it a non trivial problem to solve. We believe that by completely formalizing the notion of attribute and its treatment by the policy decision point, \ptackle paves the way to address the problem of attribute privacy. 

\appendix

\section{Monotonicity Proofs}\label{app:monotonicity}
All the following proofs have been encoded and verified in Isabelle/Isar. The definition of the three-valued logic and the corresponding operators can be found at \url{http://isg.rhul.ac.uk/~jason/isabelle/logic.thy}. 
The definition of the target and policy evaluation, together with the proofs of the following lemmas,
can be found at \url{http://isg.rhul.ac.uk/~jason/isabelle/ptacl.thy}. 

\renewcommand{\qed}{\hspace*{\fill}$\square$}

% \begin{Thm}\label{thm:strong-monotonicity}
% Let $p$ be a policy whose policy tree contains \emph{monotonic} targets $t_1,\dots,t_k$ and let $q$ be a request.
% Then for any $q' \subseteq q$, $\sem{p}(q) \subseteq \sem{p}(q')$.
% \end{Thm}
\begin{proof}
We prove the Theorem \ref{thm:strong-monotonicity} by induction over the structure of $p$.

\begin{compactenum}
	\item If $p = d$, then $\sem{p}(q) = \sem{p}(q') = d$ and we can conclude. 
        % Either $\sem{t}(q) = \sem{t}(q')$, and in this case $\sem{p}(q) = \sem{p}(q')$, and we can conclude.
        % Or $\sem{t}(q) \in \set{\nomatch, \match}$ and $\sem{t}(q') = \noreqmatch$; in this case, $\sem{p}(q) \in \set{\set{\na}, \set{d}}$ and $\sem{p}(q') = \set{\na, d}$, and we can also conclude.
	\item If $p = \pnot p'$, then $\sem{p}(q) = \lnot (\sem{p'}(q))$ and $\sem{p}(q') = \lnot (\sem{p'}(q'))$.
        Moreover, by the inductive hypothesis, we have $\sem{p'}(q) \subseteq \sem{p'}(q')$, it follows $\lnot (\sem{p'}(q)) \subseteq \lnot (\sem{p'}(q'))$, and we can conclude.
	\item If $p = \dbd p'$, then $\sem{p}(q) = \kweakening (\sem{p'}(q))$ and $\sem{p}(q') = \kweakening (\sem{p'}(q'))$.
        Moreover, by the inductive hypothesis, we have $\sem{p'}(q) \subseteq \sem{p'}(q')$, it follows $\kweakening (\sem{p'}(q)) \subseteq \kweakening (\sem{p'}(q'))$, and we can conclude.
	\item If $p =  (p_1 \pand p_2)$, then let $X_1 = \sem{p_1}(q)$, $X_2 = \sem{p_2}(q)$, $X'_1 = \sem{p_1}(q')$ and $X'_2 = \sem{p_2}(q')$, and let us show that $X_1 \strongkland X_2 \subseteq X'_1 \strongkland X'_2$.
        Indeed, by the inductive hypothesis, we have $X_1 \subseteq X'_1$ and $X_2 \subseteq X'_2$, and it follows that $\set{x_1 \strongkland x_2 : x_1 \in X_1, x_2 \in X_2} \subseteq \set{x_1 \strongkland x_2 : x_1 \in X'_1, x_2 \in X'_2}$, and we can conclude.

        \item If $p = (t, p')$, then by induction hypothesis, we have $\sem{p'}(q) \subseteq \sem{p'}(q')$.
        Since $t$ is assumed to be monotonic, four cases are possible.
  	\begin{compactenum}
  		\item Either $\sem{t}(q) = \sem{t}(q') = \nomatch$, and in this case $\sem{p}(q) = \sem{p}(q') = \set{\na}$, and we can conclude.
  		\item Or $\sem{t}(q) = \sem{t}(q') = \match$, and in this case $\sem{p}(q) = \sem{p'}(q)$ and  $\sem{p}(q') = \sem{p'}(q')$, and since we have $\sem{p'}(q) \subseteq \sem{p'}(q')$, we can conclude.
  		\item  Or $\sem{t}(q) = \sem{t}(q') = \noreqmatch$, and in this case  $\sem{p}(q) = \set{\na} \cup \sem{p'}(q)$ and  $\sem{p}(q') = \set{\na}\cup\sem{p'}(q')$, and we can similarly conclude.
  		\item Or $\sem{t}(q) \in \set{\nomatch, \match}$ and $\sem{t}(q') = \noreqmatch$, and in this case, $\sem{p}(q) \in \set{\set{\na}, \sem{p'}(q)}$ and $\sem{p}(q') = \set{\na}\cup \sem{p'}{q'}$, and we can also conclude.\qed
    \end{compactenum}
\end{compactenum}
\end{proof}

\begin{Lem}
Given any monotonic target $t$, the target $\tnot t$ is also monotonic.
\end{Lem}
\begin{proof}
Let $q, q'$ be two queries such that $q' \subseteq q$. Let us show that
$\lnot (\sem{t}(q')) \in \set{\noreqmatch, \lnot (\sem{t}(q))}$. Three cases are possibles.

\begin{compactenum}
	\item If $\lnot (\sem{t}(q)) = \match$, then by definition of $\lnot$, $\sem{t}(q) = \nomatch$. Since $t$ is monotonic, we have $\sem{t}(q') \in \set{\noreqmatch, \nomatch}$, and therefore we can conclude that $\lnot \sem{t}(q') \in \set{\noreqmatch, \match}$.
	
	\item If $\lnot (\sem{t}(q)) = \nomatch$, then by definition of $\lnot$, $\sem{t}(q) = \match$. Since $t$ is monotonic, we have $\sem{t}(q') \in \set{\noreqmatch, \match}$, and therefore we can conclude that $\lnot \sem{t}(q') \in \set{\noreqmatch, \nomatch}$.

	\item If $\lnot (\sem{t}(q)) = \noreqmatch$, then by definition of $\lnot$, $\sem{t}(q) = \noreqmatch$. Since $t$ is monotonic, we have $\sem{t}(q') = \noreqmatch$, and therefore we can conclude that $\lnot \sem{t}(q') =\noreqmatch$.\qed
\end{compactenum}

\end{proof}

\begin{Lem}
Given two monotonic targets $t_1$ and $t_2$, the target $t_1 \tor\, t_2$ is also monotonic.
\end{Lem}
\begin{proof}
Let $q, q'$ be two queries such that $q' \subseteq q$. Let us show that
$(\sem{t_1}(q') \strongklor \sem{t_2}(q')) \in \set{\noreqmatch, (\sem{t_1}(q) \strongklor \sem{t_2}(q))}$. Three cases are possibles.

\begin{compactenum}
	\item If $\sem{t_1}(q) \strongklor \sem{t_2}(q) = \match$, then by definition of $\strongklor$, $\sem{t_1}(q) = \match$ or $\sem{t_2}(q) = \match$. Since $t_1$ and $t_2$ are monotonic, we have $\sem{t_1}(q') \in \set{\noreqmatch, \match}$ or  $\sem{t_2}(q') \in \set{\noreqmatch, \match}$, and therefore we can conclude that $\sem{t_1}(q') \strongklor \sem{t_2}(q') \in \set{\noreqmatch, \match}$.
	
	\item If $\sem{t_1}(q) \strongklor \sem{t_2}(q) = \nomatch$, then by definition of $\strongklor$, $\sem{t_1}(q) = \sem{t_2}(q) = \nomatch$. % and $\sem{t_2}(q) = \nomatch$.
	Since $t_1$ and $t_2$ are monotonic, we have $\sem{t_1}(q') \in \set{\noreqmatch, \nomatch}$ and  $\sem{t_2}(q') \in \set{\noreqmatch, \nomatch}$, and  we can conclude that $\sem{t_1}(q') \strongklor \sem{t_2}(q') \in \set{\noreqmatch, \nomatch}$.
	
	\item If $\sem{t_1}(q) \strongklor \sem{t_2}(q) = \noreqmatch$, then by definition of $\strongklor$, at least one target among $\set{t_1, t_2}$ evaluates to $\noreqmatch$ while the other one evaluates either to $\nomatch$ or to $\noreqmatch$. Let us consider that $\sem{t_1}(q) = \noreqmatch$ and $\sem{t_2}(q) \in \set{\noreqmatch, \nomatch}$, the symmetrical case being equivalent. Since $t_1$ and $t_2$ are monotonic, we have $\sem{t_1}(q') = \noreqmatch$ and  $\sem{t_2}(q') \in \set{\noreqmatch, \nomatch}$, and therefore we can conclude that $\sem{t_1}(q') \strongklor \sem{t_2}(q') = \noreqmatch$.\qed
\end{compactenum}
\end{proof}

\noindent
\begin{Lem}
Given two monotonic targets $t_1$ and $t_2$, the targets $t_1 \tand_{\strongkland}\, t_2,  t_1 \tand\, t_2$ and  $t_1 \tor_{\weakklor}\, t_2$, where $\tand_{\strongkland}$ and  $\tor_{\weakklor}$ are semantically defined by the operators $\strongkland$ and $\weakklor$, respectively, are also monotonic.
\end{Lem}
\begin{proof}
The proofs follow directly from the definitions of $\strongkland, \weakkland$ and $\weakklor$ using $\strongklor$ and $\lnot$.\qed
\end{proof}

% \begin{Lem}
% Given two monotonic targets $t_1$ and $t_2$, the target $t_1 \tand\, t_2$ is also monotonic.
% \end{Lem}
% \begin{proof}
% The proof follows directly from the fact that the operator $\weakkland$	can be defined using only $\strongkland$, $\strongklor$ and $\lnot$, and from the previous lemmas.
% \end{proof}
%
% \begin{Lem}
% Given two monotonic targets $t_1$ and $t_2$, the target $t_1 \tor_{\weakklor}\, t_2$, where $\tor_{\weakklor}$ stands for the syntactic sugar corresponding to the operator $\weakklor$, is also monotonic.
% \end{Lem}
% \begin{proof}
% The proof follows directly from the fact that the operator $\weakklor$ can be defined using only $\weakkland$ and $\lnot$ and from the previous lemmas.
% \end{proof}

\section{Weak Monotonicity Proofs}
\label{app:weak-monotonicity}

% \begin{Thm}\label{thm:weak-monotonicity}
% Let $p$ be a policy defined without the operator $\dbd$ and whose policy tree contains \emph{weakly monotonic} targets $t_1,\dots,t_k$ and let $q$ be a request.
% Then for any $q' \subseteq q$, if $\sem{p}(q') = \set{d}$, with $d \in \set{\allow, \deny}$, then $\sem{p}(q) = \sem{p}(q')$.
% \end{Thm}
\begin{proof}
We prove the first case of the Theorem~\ref{thm:weak-monotonicity} by induction over the structure of $p$, the proof of the second case is similar.

\begin{compactenum}
	\item If $p = d$, it follows that $\sem{p}(q) = \sem{p}(q')$ and we can conclude. 
% from $\sem{p}(q') = \set{d}$, with $d \in \set{\allow, \deny}$,
		% that $\sem{t}(q') = \match$ and $d = d'$.
		% Since $t$ is assumed to be weakly monotonic, we can deduce that  $\sem{t}(q) = \match$, and therefore we have $\sem{p}(q) = \set{d'} = \sem{p}(q')$ and we 				
		% can conclude. 	
	\item If $p = \pnot p'$, then $\sem{p}(q') = \lnot (\sem{p'}(q'))$. From $\sem{p}(q') = \set{d}$, with $d \in \set{\allow, \deny}$, we can deduce that
	 	$\sem{p'}(q') = \set{\lnot d}$. Moreover, by the inductive hypothesis, we have $\sem{p'}(q) = \sem{p'}(q')$,
		it follows $\lnot (\sem{p'}(q)) = \set{d} = \lnot (\sem{p'}(q'))$, and we can conclude.
	\item If $p = (p_1 \pand\,p_2)$, then let $X_1 = \sem{p_1}(q)$, $X_2 = \sem{p_2}(q)$, $X'_1 = \sem{p_1}(q')$ and $X'_2 = \sem{p_2}(q')$. 
          Two cases are possible, depending on the value of $\sem{p}(q')$.
          \begin{compactenum}
          \item Either $\sem{p}(q') = \set{\allow}$, and in this case, % from the definition of $\strongkland$,
            we can deduce that $X'_1 = X'_2 = \set{\allow}$.
            % and $X'_2 = \set{\allow}$.
            By the inductive hypothesis, it follows that $X_1 = X_2 =\set{\allow}$ %and $X_2 = \set{\allow}$
            and therefore we have
            $\sem{p}(q) = \set{\allow} = \sem{p}(q')$, and we can conclude.
            
          \item Or $\sem{p}(q') = \set{\deny}$, and in this case, from the definition of $\strongkland$, we can deduce that either $X'_1 = \set{\deny}$
            or $X'_2 = \set{\deny}$. By the inductive hypothesis, it follows that either $X_1 = \set{\deny}$ or $X_2 = \set{\deny}$ and therefore we have
            $\sem{p}(q) = \set{\deny} = \sem{p}(q')$, and we can conclude.\qed
          \end{compactenum}
          
          \item If $p = (t, p')$, then from $\sem{p}(q') = \set{d}$, with $d \in \set{\allow, \deny}$, we have that $\sem{t}(q') = \match$, 
            and since $t$ is assumed to be weakly monotonic, we can deduce that  $\sem{t}(q) = \match$. It follows that 
            $\sem{p}(q) = \sem{p'}(q)$ and $\sem{p}(q') = \sem{p'}(q')$, and by induction hypothesis, we can conclude. 
	\end{compactenum}
\end{proof}

\begin{Lem}
Given any weakly monotonic target $t$, $\tweakening t$ is weakly monotonic.
\end{Lem}
\begin{proof}
Let $q, q'$ be two queries such that $q' \subseteq q$. Let us show that
$\kweakening (\sem{t}(q')) \preccurlyeq \kweakening (\sem{t}(q))$. Since $\kweakening (\sem{t}(q)) \neq \noreqmatch$, by definition of $\kweakening$, only
	two cases are possibles.

\begin{compactenum}
	\item Either $\kweakening (\sem{t}(q)) = \match$% , then by definition of $\preccurlyeq$,
	and we can trivially conclude.
	
	\item Or $\kweakening (\sem{t}(q)) = \nomatch$, then by definition of $\kweakening$, $\sem{t}(q) \in \set{\noreqmatch, \nomatch}$.
	Since $t$ is weakly monotonic, we have $\sem{t}(q') \preccurlyeq \sem{t}(q')$ and thus $ \sem{t}(q') \in \set{\noreqmatch, \nomatch}$. It follows that
	$ \kweakening (\sem{t}(q')) = \nomatch$, and we can conclude.\qed
\end{compactenum}

\end{proof}

\begin{Lem}
Given two weakly monotonic targets $t_1$ and $t_2$, the target $t_1 \tor\, t_2$ is also weakly monotonic.
\end{Lem}
\begin{proof}
Let $q, q'$ be two queries such that $q' \subseteq q$. Let us show that
$(\sem{t_1}(q') \strongklor \sem{t_2}(q')) \preccurlyeq (\sem{t_1}(q) \strongklor \sem{t_2}(q))$. Three cases are possibles.

\begin{compactenum}
	\item If $\sem{t_1}(q) \strongklor \sem{t_2}(q) = \match$, then % by definition of $\preccurlyeq$
	we can trivially conclude.
	
	\item If $\sem{t_1}(q) \strongklor \sem{t_2}(q) = \nomatch$, then by definition of $\strongklor$, $\sem{t_1}(q) = \nomatch$ and $\sem{t_2}(q) = \nomatch$.
	Since $t_1$ and $t_2$ are weakly monotonic, we have $\sem{t_1}(q') \in \set{\noreqmatch, \nomatch}$ and  $\sem{t_2}(q') \in \set{\noreqmatch, \nomatch}$.
	By definition of $\strongklor$, it follows $\sem{t_1}(q') \strongklor \sem{t_2}(q') \in \set{\noreqmatch, \nomatch}$ and we can conclude.
	
	\item If $\sem{t_1}(q) \strongklor \sem{t_2}(q) = \noreqmatch$, then by definition of $\strongklor$, at least one target among $\set{t_1, t_2}$ evaluates
	to $\noreqmatch$ while the other one evaluates either to $\nomatch$ or to $\noreqmatch$. Let us consider that $\sem{t_1}(q) = \noreqmatch$
	and $\sem{t_2}(q) \in \set{\noreqmatch, \nomatch}$, the symmetrical case being equivalent. Since $t_1$ and $t_2$ are weakly monotonic,
	we have $\sem{t_1}(q') = \noreqmatch$ and  $\sem{t_2}(q') \in \set{\noreqmatch, \nomatch}$. By definition of $\strongklor$, it follows
	$\sem{t_1}(q') \strongklor \sem{t_2}(q') = \noreqmatch$, and we can conclude.\qed
\end{compactenum}
\end{proof}

\begin{Lem}\label{lem:wm-tand}
Given two weakly monotonic targets $t_1$ and $t_2$, the target $t_1 \tand\, t_2$ is also weakly monotonic.
\end{Lem}
\begin{proof}
Let $q, q'$ be two queries such that $q' \subseteq q$. Let us show that
$(\sem{t_1}(q') \weakkland \sem{t_2}(q')) \preccurlyeq (\sem{t_1}(q) \weakkland \sem{t_2}(q))$. Three cases are possibles.

\begin{compactenum}
	\item If $\sem{t_1}(q) \weakkland \sem{t_2}(q) = \match$, then % by definition of $\preccurlyeq$
	we can trivially conclude.
	
	\item If $\sem{t_1}(q) \weakkland \sem{t_2}(q) = \nomatch$, then by definition of $\weakkland$, either $\sem{t_1}(q) = \nomatch$ or $\sem{t_2}(q) = \nomatch$.
	Since $t_1$ and $t_2$ are weakly monotonic, we have either $\sem{t_1}(q') \in \set{\noreqmatch, \nomatch}$ or $\sem{t_2}(q') \in \set{\noreqmatch, \nomatch}$.
	By definition of $\weakkland$, it follows $\sem{t_1}(q') \weakkland \sem{t_2}(q') \in \set{\noreqmatch, \nomatch}$ and we can conclude.
	
	\item If $\sem{t_1}(q) \weakkland \sem{t_2}(q) = \noreqmatch$, then by definition of $\weakkland$, either $\sem{t_1}(q) = \noreqmatch$ or
	$\sem{t_2}(q) = \noreqmatch$.
	Since $t_1$ and $t_2$ are weakly monotonic, we have either $\sem{t_1}(q') = \noreqmatch$ or $\sem{t_2}(q') = \noreqmatch$,
	and it follows $\sem{t_1}(q') \weakkland \sem{t_2}(q') = \noreqmatch$, allowing us to conclude.\qed
\end{compactenum}
\end{proof}

\begin{Lem}
Given two weakly monotonic targets $t_1$ and $t_2$, the target $t_1 \tor_{\weakklor}\, t_2$, where $\tor_{\weakklor}$ stands for the syntactic sugar corresponding to the operator $\weakklor$, is also weakly monotonic.
\end{Lem}
\begin{proof}
% Similar to the proof of Lemma~\ref{lem:wm-tand}.
Let $q, q'$ be two queries such that $q' \subseteq q$. Let us show that
$(\sem{t_1}(q') \weakklor \sem{t_2}(q')) \preccurlyeq (\sem{t_1}(q) \weakklor \sem{t_2}(q))$. Three cases are possibles.

\begin{compactenum}
	\item If $\sem{t_1}(q) \weakklor \sem{t_2}(q) = \match$, then % by definition of $\preccurlyeq$
	we can trivially conclude.
	
	\item If $\sem{t_1}(q) \weakklor \sem{t_2}(q) = \nomatch$, then % by definition of $\weakklor$, we have
	$\sem{t_1}(q) = \sem{t_2}(q) =\nomatch$.
	Since $t_1$ and $t_2$ are weakly monotonic, we have $\sem{t_1}(q') \in \set{\noreqmatch, \nomatch}$ and $\sem{t_2}(q') \in \set{\noreqmatch, \nomatch}$.
	By definition of $\weakklor$, it follows $\sem{t_1}(q') \weakklor \sem{t_2}(q') \in \set{\noreqmatch, \nomatch}$ and we can conclude.
	
	\item If $\sem{t_1}(q) \weakklor \sem{t_2}(q) = \noreqmatch$, then by definition of $\weakklor$, either $\sem{t_1}(q) = \noreqmatch$ or
	$\sem{t_2}(q) = \noreqmatch$.
	Since $t_1$ and $t_2$ are weakly monotonic, we have either $\sem{t_1}(q') = \noreqmatch$ or $\sem{t_2}(q') = \noreqmatch$.
	By definition of $\weakklor$, it follows $\sem{t_1}(q') \weakklor \sem{t_2}(q') = \noreqmatch$ and we can conclude.\qed
\end{compactenum}
\end{proof}

\end{document}